	\documentclass[journal]{IEEEtran}
	\usepackage{graphicx,amssymb,amsmath}
	\usepackage{multicol}
	\usepackage[noadjust]{cite}
	\usepackage{setspace}
	\usepackage{subfigure}
	\usepackage{graphicx}
	\usepackage{float}
	\usepackage {url}
	\usepackage{stfloats}
	\usepackage{amsthm,pifont}
	\usepackage{flushend}
	\usepackage{cases,subeqnarray}
	\usepackage{bm,multirow,bigstrut}
	\usepackage{amsmath, amsthm, amssymb}
	\usepackage{textcomp}
	\usepackage{latexsym,bm}
	\usepackage{booktabs}
	\usepackage{xcolor}
	\usepackage{mathtools}
	\usepackage{dsfont}
	\usepackage{extarrows}
	\usepackage{epsfig}
	\usepackage{epsfig}
	\usepackage{epstopdf}
	\usepackage{subfigure}	
	\theoremstyle{plain}

	\theoremstyle{plain}
	\newtheorem{rem}{Remark}
	\newtheorem{them}{Theorem}
	
	\newtheorem{prop}{Proposition}

	\IEEEoverridecommandlockouts
	
\begin{document}
	\title{Physical Layer Security Enhancement With Reconfigurable Intelligent Surface-Aided Networks}
	\author{Jiayi~Zhang,~\IEEEmembership{Senior Member,~IEEE}, Hongyang Du, Qiang Sun, Bo Ai,~\IEEEmembership{Senior~Member,~IEEE}, \\and Derrick Wing Kwan Ng,~\IEEEmembership{Fellow,~IEEE}
	\thanks{J.~Zhang and Q. Sun are with the School of Information Science and Technology, Nantong University,
	Nantong 226019, China. J.~Zhang is also with with the School of Electronic and Information Engineering, Beijing Jiaotong University, Beijing 100044, P. R. China. (e-mail: sunqiang@ntu.edu.cn).}
	\thanks{H. Du is with the School of Electronic and Information Engineering, Beijing Jiaotong University, Beijing 100044, P. R. China. (e-mail: 17211140@bjtu.edu.cn)}	
	\thanks{B. Ai is with the State Key Laboratory of Rail Traffic Control and Safety, Beijing Jiaotong University, Beijing 100044, China (e-mail: boai@bjtu.edu.cn).}
	\thanks{D. W. K. Ng is with the School of Electrical Engineering and Telecommunications, University of New South Wales, NSW 2052, Australia. (e-mail: w.k.ng@unsw.edu.au).}
	}
	\maketitle
	\vspace{-1cm}
	\begin{abstract}
	Reconfigurable intelligent surface (RIS)-aided wireless communications have drawn significant attention recently. We study the physical layer security of the downlink RIS-aided transmission framework for randomly located users in the presence of a multi-antenna eavesdropper. To show the advantages of RIS-aided networks, we consider two practical scenarios: Communication with and without RIS. In both cases, we apply the stochastic geometry theory to derive exact probability density function (PDF) and cumulative distribution function (CDF) of the received signal-to-interference-plus-noise ratio. Furthermore, the obtained PDF and CDF are exploited to evaluate important security performance of wireless communication including the secrecy outage probability, the probability of nonzero secrecy capacity, and the average secrecy rate. Monte-Carlo simulations are subsequently conducted to validate the accuracy of our analytical results. {\color{blue}Compared with traditional MIMO systems, the RIS-aided system offers better performance in terms of physical layer security. In particular, the security performance is improved significantly by increasing the number of reflecting elements equipped in a RIS. However, adopting RIS equipped with a small number of reflecting elements cannot improve the system performance when the path loss of NLoS is small.}
	\end{abstract}
	\IEEEpeerreviewmaketitle
	\section{Introduction}
	Recently, reconfigurable intelligent surface (RIS) has been proposed as a promising technique as it can achieve high spectral-/energy-efficiency through adaptively controlling the wireless signal propagation environment \cite{cui2014coding}. Specifically, RIS is a planar array which comprises a large number of nearly passive reflecting elements. By equipping the RIS with a controller, each element of the RIS can independently introduce a phase shift on the reflected signal. Besides, a RIS can be easily coated on existing infrastructures such as walls of buildings, facilitating low-cost and low-complexity implementation. By smartly adjusting the phase shifts induced by all the reflecting elements, {\color{blue}the RIS can bring} various potential advantages to the system such as enriching the channel by deliberately introducing more multi-paths, increasing the coverage area, or beamforming, while consuming very low amount of energy due to the passive nature of its elements \cite{yu2021smart}.
	Recently, RIS has been introduced into many wireless communication systems. For instance, the authors in \cite{wu2019intelligent} studied a RIS-aided single-user multiple-input single-output (MISO) system and optimized the introduced phase shifts to maximize the total received signal power at the users. In \cite{huang2019reconfigurable}, various designs were proposed for both the transmit power allocation at a base station (BS) and the phases introduced by the RIS elements to maximize the energy and the spectral efficiency of a RIS-aided multi-user MISO system. Besides, the authors of \cite{nadeem2019large} considered a downlink (DL) multiuser communication system where the signal-to-interference-plus noise ratio (SINR) was maximized for given phase shifts introduced by a RIS.
	
	In practice, communication security is always a fundamental problem in wireless networks due to its broadcast nature. Besides traditional encryption methods adopted in the application layer, physical layer security (PLS) \cite{chen2016survey,wyner1975wire,du2020on} serves as an alternative for providing secure communication in fast access fifth-generation (5G) networks. Furthermore, PLS can achieve high-quality safety performance without requiring actual key distribution, which is a perfect match with the requirements of 5G. To fully exploit the advantages of 5G, multi-antenna technology has become a powerful tool for enhancing the PLS in wireless fading networks, e.g.  \cite{wang2016physical,hong2013enhancing,goel2008guaranteeing,zheng2014transmission,geraci2014physical,wang2016artificial,zhou2010secure}. In particular, with the degrees of freedom provided by multiple antennas, a transmitter can steer its beamforming direction to exploit the maximum directivity gain to reduce the potential of signal leakage to eavesdroppers. However, there are only a few works considering the PLS in the emerging RIS-based communication systems, despite its great importance for modern wireless systems. For example, a RIS-aided secure communication system was investigated in \cite{yu2019enabling,cui2019secure} but their communication systems only consist of a transmitter, one legitimate receiver, single eavesdropper, and a RIS with limited practical applications. Furthermore, authors in \cite{chen2019intelligent} studied a DL MISO broadcast system where a BS transmits independent data streams to multiple legitimate receivers securely in the existence of multiple eavesdroppers. In order to conceive a practical RIS framework, user positions have to be taken into account using stochastic geometry for analyzing the system performance. In fact, stochastic geometry is an efficient mathematical tool for capturing the topological randomness of networks \cite{chiu2013stochastic,haenggi2012stochastic}. However, there are only a few works studying the impact of user-locations on the security performance.
	
	{\color{blue}Motivated by the aforementioned reasons, in this paper, we study the security performance of a RIS-aided DL multiple-input multiple-output (MIMO) system for randomly located roaming multi-antenna users in the presence of a multi-antenna eavesdropper. Moreover, we aim to answer a fundamental question \textit{``How much improvement can RIS bring to physical} \textit{layer security in wireless communications?''.}} The main contributions of this paper are summarized as follows:
	{\color{blue}\begin{itemize}
			\item To show the advantages of exploiting RIS to enhance the PLS in wireless communications, we consider two practical scenarios: RIS is adopted and its absence. For each case, we derive the exact closed-form expressions of SINR and its PDF and CDF, respectively. Furthermore, considering a general set-up when the direct links between the BS and the RIS-aided users exist, we obtain the CDF and outage probability (OP) expressions.
			\item Exploiting the tools from stochastic geometry, we propose a novel PLS analysis framework of RIS-aided communication systems. Novel expressions are derived to characterize the PLS performance, namely the secrecy outage probability (SOP), the probability of nonzero secrecy capacity (PNSC), and the average secrecy rate (ASR). The derived results can provide insights, i.e., the security performance is improved significantly by increasing the number of reflecting elements.
			\item We derive highly accurate and simplified closed-form approximations of the CDF expressions of SINR for both scenarios. Furthermore, we present an asymptotic OP analysis in the high-SNR regime. The derived results show that adopting RIS equipped with a small number of reflecting elements cannot improve the system performance if the path loss of NLoS is small.
	\end{itemize}}
	
	The remainder of the paper is organized as follows. In Section \ref{sec2}, we introduce the system model of the RIS-aided DL MIMO communication system and derive the exact closed-form of PDF and CDF expressions of SINR for users and the eavesdropper. Section \ref{sec3} formulates the security performance analysis in two scenarios and present the corresponding exact expressions of the performance. {\color{blue}In Section \ref{secapp}, we derive the approximations to the statistics of SINR in two scenarios, and analyze the OP in high-SINR regime. Section \ref{apofpa} presents the CDF and OP of scenario 2 when the direct links exist between BS and RIS-aided users.} Section \ref{sec4} shows the simulation results and the accuracy of the obtained expressions is validated via Monte-Carlo simulations. Finally, Section \ref{conclusion} concludes this paper.
	
	\emph{Mathematical notations}: A list of mathematical symbols and functions most frequently used in this paper is available in Table \ref{symbol}. 
	
	\begin{table}[htbp]\label{symbol}
	\caption{MATHEMATICAL SYMBOLS AND FUNCTIONS}
	\centering
	\begin{tabular}{l|p{6.5cm}}
	\toprule
	$\upsilon $ &{\textcolor{blue}{A subscript which denotes different links. $\upsilon=u_1, \cdots, u_N, u_e, u_R$ denotes the signal is sent by the BS, and destinations are $N$ users, the eavesdropper, and the RIS, respectively. $\upsilon=u_R-u_1, \cdots, u_R-u_N, u_R-u_e$ denotes the signal is reflected by the RIS, and the destinations are $N$ users and the eavesdropper, respectively. }}\\
	$\sigma$ &{\textcolor{blue}{A subscript which denotes different kinds of path loss. $\sigma={\rm L}$ denotes the LoS links, $\sigma={\rm NL}$ denotes the NLoS links.}}\\
	$\kappa $ & A subscript which denotes different antennas.\\
	${\mathbb{E}}[\cdot]$ & The mathematical expectation.\\
	$r_2$ & The radius of the disc. \\
	${\bf H}^{\rm H}$ & Conjugate transpose.  \\
	${\bf H}^{\rm T}$ & Transpose.  \\
	$ {{\mathbb C}^{x \times y}} $ & The space of $x \times y$ complex-valued matrices. \\
	$\|\cdot\|_{F}$ & The Frobenius norm.\\
	$j$ & $j = \sqrt{-1} $  \\
	$\left\{ {{\Delta _i}} \right\}_{i = 1}^L$ & $\left\{ {{\Delta _i}} \right\}_{i = 1}^L$ denotes ${\Delta _1},{\Delta _2}, \cdots ,{\Delta _L}$.\\
	$\Gamma \left( \cdot \right) $ &  Gamma function \cite[eq. (8.310/1)]{gradshteyn2007}   \\
	${B\left( { \cdot , \cdot } \right)}$ &  Beta function \cite[Eq. (8.384.1)]{gradshteyn2007}  \\
	${}_2{F_1}\!\left( { \!\cdot , \cdot ; \cdot ; \cdot \!} \right)$ &  Hypergeometric function \cite[Eq. (9.111)]{gradshteyn2007}  \\
	$G \, \substack{ m , n \\ p , q}(\cdot)$ &  Meijer's $G$-function \cite[eq. (9.301)]{gradshteyn2007}  \\
	\bottomrule
	\end{tabular}
	\end{table}
	\section{System Model And Preliminaries}\label{sec2}
	\subsection{System Description}
	\begin{figure}[t]
	\centering
	\includegraphics[scale=0.46]{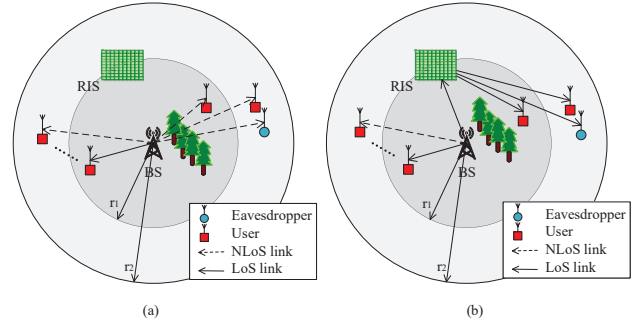}
	\caption{Two practical scenarios: Communication without (left hand side) and with RIS (right hand side).}
	\label{model1}
	\end{figure}
	
	Let us consider the DL MIMO-RIS system and assume that a BS equipped with $M$ transmit antennas (TAs) communicates with $N$ users each of which is equipped with $K$ receive antennas (RAs). a RIS is installed between the BS and the users for assisting end-to-end communication. {\color{blue}By jointly optimizing the transmit precoding at the BS and the reflect phase shifts at the RIS according to the wireless channel conditions, the reflected signals can be added constructively at the desired receiver to enhance the received signal power, which can potentially yield a secure transmission \cite{yu2020robust}.} In addition, a malicious eavesdropper who is equipped with $K$ RAs aims to eavesdrop the information of the desired user. We assume that the number of reflecting elements on the RIS is $L$. The RIS is equipped with a controller to coordinate between the BS and the RIS for both channel acquisition and data transmission \cite{wu2019intelligent}. As such, the signal received from the BS can be manipulated by the RIS via adjusting the phase shifts and amplitude coefficients of the RIS elements.
	
	On the other hand, various practical approaches have been proposed in the literature for the channel estimation of RIS-aided links \cite{wang2020channel,araujo2021channel,liu2021deep}. Thus, we can assume that the global channel state information (CSI) of users is perfectly known for joint design of beamforming \cite{mukherjee2014principles}. {\color{blue}The CSI of the eavesdropper is typically unavailable as the eavesdropper is passive to hide its existence. Therefore, the passive eavesdropping scenario is considered \cite{yang2020secrecy}. To quantify the performance of such a RIS-aided communication system, we adopt SOP, PNSC, and ASR as performance metrics of interested.}
	
	In particularly, we assume that the users are located on a disc with a radius $r_2$ according to homogeneous Poisson point processes (HPPP) \cite{haenggi2009stochastic} with density $\lambda$ and the eavesdropper will try to choose a position that is close to the legitimate user.
	
	In practice, the direct transmission link between the BS and the users may be blocked by trees or buildings. Such assumption is applicable for 5G and mmWave communication systems that are known to suffer from high path and penetration losses resulting in signal blockages. In order to unveil the benefits in adopting the RIS, we consider the following two practice of communication scenarios:
	\begin{itemize}
	\item As shown in Fig. \ref{model1} (a), for the first scenario, RIS is not adopted for enhancing the PLS and the quality of wireless communication. {\color{blue}Specifically, both LoS and NLoS links exist in our system. Without loss of generality, we focus our attention on user $n$. We use $d_{\upsilon,\sigma}$ to denote the distance between the BS and user $n$ $(n=1,\cdots,N)$. According to Table \ref{symbol}, we have $\upsilon=u_n$, $\sigma={\rm L}$ for LoS links and $\upsilon=u_n$, $\sigma={\rm NL}$ for NLoS links.}
	\item {\color{blue}For the second scenario, a RIS is deployed to leverage the LoS components with respect to both the BS and the users to assist their end-to-end communication of Fig. \ref{model1} (b). Thus, the NLoS users in scenario 1 now can communicate with BS through two LoS links with the help of RIS. Moreover, the locations of the BS and the RIS are fixed, hence we assume that the distance between the BS and the RIS is known and denoted by $d_{u_R,{\rm L}}$. In addition, the distance between the RIS and user $n$ is denoted by $d_{u_R-u_n,{\rm L}}$.}
	\end{itemize}
	\subsection{Blockage Model}
	A blockage model was proposed in \cite{singh2015tractable}, which can be regarded as an accurate approximation of the statistical blockage model \cite{di2015stochastic} and incorporates the LoS ball model proposed in \cite{bai2014coverage} as a special case. In the considered system model, we adopt the blockage model to divide the users process in the spherical region around the BS into two independent HPPPs: LoS users process and NLoS users process. In particular, we define ${q_L}(r)$ as the probability that a link of length $r$ is LoS. Each access link of separation $r$ is assumed to be LoS with probability $B_1$ if $r \le r_1$ and $0$ otherwise:
	\begin{equation}\label{b1}
	{q_L}(r) = \left\{ {\begin{array}{*{20}{l}}
	{B_1,}&{{\rm{if}}\quad r \le r_1}\\
	{0,}&{{\rm{ otherwise }}}
	\end{array}} \right.,
	\end{equation}
	{\color{blue}{where $0 \le B_1 \le 1$. The parameter $B_1$ can be interpreted as the average LoS area in a circular ball with a radius of $r_1$ around the BS.}}
	
	\subsection{User Model}
	Let us assume that the users are located according to a HPPP within the disc shown as Fig. \ref{model1}. The PDF of the user locations have been derived as \cite[eq. (30)]{hou2019mimo}. As a result, the CDF of the user locations is given by
	\begin{equation}\label{cdflocation}
	{F_R}(r) = \frac{{{r^2}}}{{\left( {{r_2}^2 - r_0^2} \right)}} - \frac{{r_0^2}}{{\left( {{r_2}^2 - r_0^2} \right)}}, \qquad{\rm{ if }}\quad{r_0} < r < R,
	\end{equation}
	{\textcolor{blue}{where $r_0$ is the minimum distance which is used to avoid encountering a singularity \cite{ding2015application}}}. Thus, the probability of the distance between the user and the BS is less than $r_1$ can be expressed as
	\begin{equation}\label{b2}
	{B_2} \triangleq {\rm{Pr}}\left( {r \le {r_1}} \right) = {F_R}({r_1}) = \frac{{r_1^2}}{{\left( {{r_2}^2 - r_0^2} \right)}} - \frac{{r_0^2}}{{\left( {{r_2}^2 - r_0^2} \right)}}.
	\end{equation}
	
	\subsection{Path Loss Model}
	{\color{blue}
	For the first communication scenario without a RIS, different path loss equations are applied to model the LoS and NLoS links as \cite{singh2015tractable,bai2014coverage}
	
	\begin{equation}
	L\left( d \right){=}\left\{ {\begin{array}{*{20}{c}}\!\!\!
	{{d_{u_n,{\rm L}}^{ - {\alpha _1}}}},\!\!\!\\\!\!\!
	{{d_{u_n,{\rm NL}}^{ - {\alpha _2}}}},\!\!\!\!
	\end{array}} \right.\begin{array}{*{20}{c}}\!\!\!\!\!
	{{\rm{if\: BS}} \to {\rm{user \:{\it n}\: link\: is\: LoS\: link.}}}\\
	{{\rm{if\: BS}} \to {\rm{user \:{\it n}\: link\: is\: NLoS\: link,}}}
	\end{array}
	\end{equation}
	where ${\alpha _1}$ and ${\alpha _2}$ are the LoS and NLoS path loss exponents, respectively. Typical values of ${\alpha _1}$ and ${\alpha _2}$ are defined in \cite[Table 1]{akdeniz2014millimeter}, while ${\alpha _1} < {\alpha _2}$ hold in general.
	
	For the second communication scenario with the assistance by the RIS, recently, the free-space path loss models of RIS-aided wireless communications are developed for different situations in \cite[Proposition 1]{tang2019wireless}. Authors in \cite{tang2019wireless} proposed that the free-space path loss of RIS-aided communications is proportional to ${(d_{u_R,{\rm L}}d_{u_R-u_n,{\rm L}})}^{ 2}$ in the far field case. Thus, we obtain
	\begin{equation}
	L \left( {{d_{u_R,{\rm L}}},{d_{u_R-u_n,{\rm L}}}} \right) = C_{L_1}C_{L_2}d_{u_R,{\rm L}}^{ - 2}d_{u_R-u_n,{\rm L}}^{ - 2},
	\end{equation}
	where $C_{L_1}$ and $C_{L_2}$ denote the path loss intercepts of BS-RIS and RIS-user links, respectively.}
	
	\subsection{Small-Scale Fading}
	To unify the system performance with different channel environment, generalized fading distributions have been proposed that include the most common fading distributions as special cases \cite{nauryzbayev2018on,rabie2019full,du2020on}. It has been shown recently that the Fisher-Snedecor $\mathcal{F}$ composite fading model can provide a more comprehensive modeling and characterization of the simultaneous occurrence of multipath fading and shadowing, which is generally more accurate than most other generalized fading models \cite{yoo2017the}. Furthermore, the Fisher-Snedecor $\mathcal{F}$ model is more general and includes several fading distributions as special cases, while being more mathematically tractable. For example, authors in \cite{hou2019mimo} conceived a RIS-aided MIMO framework using Nakagami-$m$ distribution. Note that Fisher-Snedecor $\mathcal{F}$ distribution includes the case of Nakagami-$m$ distribution for $m_s \to \infty$ as one of its subsequent special cases, such as Rayleigh $(m = 1)$ and one-sided Gaussian $(m = 1/2)$. Motivated by this, the Fisher-Snedecor $\mathcal{F}$ distribution has been adopted to analyze the RIS-aided systems, i.e., RIS-aided Internet-of-Things networks \cite{makarfi2020reconfigurable}.
	
	To conceive a practical RIS framework, we assume that the small-scale fading of each link follows Fisher-Snedecor $\mathcal{F}$ fading distributions \cite{makarfi2020reconfigurable}. {\color{blue}The channel correlation between RIS elements may exist because the electrical size of RIS's reflecting elements is between $\lambda/8$ and $\lambda/4$ in principle, where $\lambda$ is a wavelength of the signal \cite{scui2017information}. However, it is hard to model the correlation in RIS because a tractable model for capturing such unique characteristics has not been reported in the literature yet. Therefore, we consider a scenario where the correlation is weak enough to be ignored, i.e., the electrical size of RIS's reflecting elements is larger than $\lambda/4$ because the frequency of signals is high \cite{dardari2019communicating}, \cite{bjornson2020power}. Thus, we can assume that the small scale fading components are independent of each other.}
	
	For the first communication scenario, in order to characterize the LoS and NLoS links between the BS and users (eavesdropper), the small-scale fading matrices are defined as
	\begin{equation}\label{case1}
	{{{\bf Q}}_{\upsilon,\sigma }} = \left[ {\begin{array}{*{20}{c}}
	{{q_{1,1}^{\upsilon,\sigma }}}& \cdots &{{q_{1,M}^{\upsilon,\sigma }}}\\
	\vdots & \vdots & \vdots \\
	{{q_{K,1}^{\upsilon,\sigma }}}& \cdots &{{q_{K,M}^{\upsilon,\sigma }}}
	\end{array}} \right],
	\end{equation}
	where ${{{\bf Q}}_{\upsilon,\sigma }}$ $(\sigma={\rm L}, {\rm NL})$ is a $K \times M$ matrix. Letting $\upsilon=u_n$, we obtain the small-scale fading matrix between the BS and user $n$. Using \cite[Eq. (5)]{yoo2019comprehensive}, the PDF of the elements of \eqref{case1} can be expressed as
	\begin{equation}\label{FisherPDF}
	{f_X}(x) = \frac{{2{m^m}{{\left( {\left( {{m_s} - 1} \right)\Omega } \right)}^{{m_s}}}}{{x^{2m - 1}}}}{{B\left( {m,{m_s}} \right)}{{{\left( {m{r^2} + \left( {{m_s} - 1} \right)\Omega } \right)}^{m + {m_s}}}}},
	\end{equation}
	where $\Omega=\mathbb{E}\left[r^{2}\right]$ is the mean power, $m$ and $m_s$ are physical parameters which represent the fading severity and shadowing parameters, respectively.
	
	For the second case, the small-scale fading matrices for the LoS links between the BS and users (eavesdropper) are same as the first case. In addition, in order to model the LoS links between the BS and the RIS, the small-scale fading matrix is defined as
	\begin{equation}
	{{{\bf Q}}_{u_R,{\rm L}}} = \left[ {\begin{array}{*{20}{c}}
	{{q_{1,1}^{u_R,{\rm L}}}}& \cdots &{{q_{1,M}^{u_R,{\rm L}}}}\\
	\vdots & \vdots & \vdots \\
	{{q_{L,1}^{u_R,{\rm L}}}}& \cdots &{{q_{L,M}^{u_R,{\rm L}}}}
	\end{array}} \right],
	\end{equation}
	where ${{{\bf Q}}_{u_R,{\rm L}}}$ is a $L \times M$ matrix.
	
	In order to model the LoS links between the RIS and users (eavesdropper), the small-scale fading matrices are defined as
	\begin{equation}
	{{{\bf Q}}_{\upsilon,{\rm L}}} = \left[ {\begin{array}{*{20}{c}}
	{{q_{1,1}^{\upsilon,{\rm L}}}}& \cdots &{{q_{1,L}^{\upsilon,{\rm L}}}}\\
	\vdots & \vdots & \vdots \\
	{{q_{K,1}^{\upsilon,{\rm L}}}}& \cdots &{{q_{K,L}^{\upsilon,{\rm L}}}}
	\end{array}} \right],
	\end{equation}
	where ${{{\bf Q}}_{\upsilon,{\rm L}}}$ is a $K \times L$ matrix. Letting $\upsilon=u_R-u_n$, we get the small-scale fading matrix between the RIS and user $n$.
	
	\subsection{Directional Beamforming}
	Highly directional beamforming antenna arrays are deployed at the BS to compensate the significant path-loss in the considered system. For mathematical tractability and similar to \cite{singh2015tractable,di2015stochastic}, the antenna pattern of users can be approximated by a sectored antenna model in \cite{wang2010capacity} which is given by
	{\color{blue}\begin{equation}\label{antennasgain}
	{\cal G}_{\upsilon, \kappa} (\theta ) = \left\{ {\begin{array}{*{20}{l}}
	{{{ G}_{\upsilon, \kappa}}},&{{\rm{ if }}\:\:|\theta | \le {\theta _c}},\\
	{{{ g}_{\upsilon, \kappa}}},&{{\rm{ otherwise }}},
	\end{array}} \right. \quad (\kappa=1,\cdots,K),
	\end{equation}
	where ${\theta}$, distributed in $\left[ {0,2\pi } \right]$, is the angle between the BS and the user}, $\theta _c$ denotes the beamwidth of the main lobe, for each TA, ${{{ G}_{\upsilon, \kappa}}}$ and ${{{ g}_{\upsilon, \kappa}}}$ are respectively the array gains of main and sidelobes. In practice, the BS can adjust their antennas according to the CSI. In the following, we denote the boresight direction of the antennas as ${0^ \circ }$. For simplifying the performance analysis, different antennas of the authorized users and the malicious eavesdropper are assumed to have same array gains. Thus, without loss of generality, we assume that {\color{blue}${\cal G}_{\upsilon, \kappa}={\cal G}_{\upsilon}$}, ${ G}_{\upsilon, \kappa}={ G}_{\upsilon}$ and ${g}_{\upsilon, \kappa}={ g}_{\upsilon}$.
	
	\subsection{SINR Analysis}
	Let us consider a composite channel model of large-scale and small-scale fading. It is assumed that the distance ${d_{\upsilon, \sigma }}$, $\upsilon=u_1,\cdots,u_N,u_e,$ and $\sigma={\rm L}, {\rm NL}$ are independent but not identically distributed (i.n.i.d.) and the large-scale fading is represented by the path loss. In the DL transmission, the complex baseband transmitted signal at the BS can be then expressed as
	\begin{equation}
	x = \sum\limits_{n = 1}^N {{{\bf p}_{u_n}}{s_{u_n}}},
	\end{equation}
	{\color{blue}where ${s_{u_n}}$ $(n=1,\cdots,N)$ are i.i.d. random variables (RVs) with zero mean and unit variance, denoting the information-bearing symbols of users, and ${\bf{p}}_{u_n}\in {\mathbb C}^{M\times1}$ is the corresponding beamforming vector.} The transmit power consumed at the BS can be expressed as
	\begin{equation}
	P = \sum\limits_{n = 1}^N {{{\left\| {{{\bf p}_{u_n}}} \right\|}^{{2}}}}.
	\end{equation}	
	For the first communication scenario, {\color{blue}e.g., Fig. \ref{model1} (a)}, there are both LoS and NLoS links. The signal received by user $n$ or the eavesdropper from the BS can be expressed as
	
	{\color{blue}\begin{align}\label{signal1}
	y_{\upsilon,\sigma}  =&\sqrt {{G_\upsilon }L\left( {{d_{\upsilon ,\sigma }}} \right)} {{\bf{v}} _\upsilon } {{\bf{Q}}_{\upsilon ,\sigma }}{{\bf p}_\upsilon }{s_\upsilon }\notag\\
	&+ \underbrace {\sqrt {{G_\upsilon }L\left( {{d_{\upsilon ,\sigma }}} \right)}\sum\limits_{i \ne \upsilon }^{N} {{{\bf{v}} _\upsilon} {{\bf{Q}}_{\upsilon ,\sigma }}{\bf p}_i{s_i}} }_{{\rm{Multi-user\: interference}}} + {{\bf{v}} _\upsilon }{{\bf N}_0},
	\end{align}
	where $\upsilon = u_n, u_e$, $\sigma={\rm L}, {\rm NL}$, ${\bf{v}}_{u_n}\in {\mathbb C}^{1\times K}$ is the detection vector of user $n$, and ${\bf N}_0\in {\mathbb C}^{K\times1}$ denotes the additive white Gaussian noise, which is modeled as a realization of a zero-mean complex circularly symmetric Gaussian variable with variance $\sigma_N^{2}$.}
	
	For the second communication scenario, {\color{blue}e.g., Fig. \ref{model1} (b),} there are LoS and RIS-aided links. For LoS links, the signal received by the user or the eavesdropper can be obtained by letting $\sigma={\rm L}$ in \eqref{signal1}. For RIS-aided links, the signal can be expressed as
	{\color{blue}\begin{align}
	&y_{{\rm{RIS}}}^\upsilon  =\sqrt {{G_\upsilon }{L\left( {{d_{u_R,{\rm L}}},{d_{u_R-\upsilon,{\rm L}}}} \right)}}{{\bf{v}} _\upsilon } {{\bf{Q}}_{u_R-\upsilon,{\rm L}}}{\bf{\Phi }}{{\bf{Q}}_{u_R,{\rm L}}} {{\bf p}_\upsilon }{s_\upsilon } \notag\\
	&+\!{{\bf{v}} _\upsilon }{{\bf N}_0}\!+\!\! \underbrace {\sqrt {{G_\upsilon }{L\!\left(\! {{d_{u_R,{\rm L}}},{d_{u_R\!-\!\upsilon,{\rm L}}}} \!\right)}}\!\sum\limits_{i \ne \upsilon }^{N}\!{ {{\bf{v}} _\upsilon }{{\bf{Q}}_{u_R\!-\!\upsilon,{\rm L}}}{\bf{\Phi }}{{\bf{Q}}_{u_R,{\rm L}}}  {{\bf p}_i}{s_i}} }_{{\rm{Multi-user\: interference}}},
	\end{align}}
	where $\upsilon = u_n, u_e$, ${\bf{\Phi }} \buildrel \Delta \over = {\mathop{\rm diag}\nolimits} \left[ {{\beta _1}{\phi _1},{\beta _2}{\phi _2}, \cdots ,{\beta _{L}}{\phi _{L}}} \right]$ is a diagonal matrix accounting for the effective phase shift introduced by all the elements of the RIS \footnote{{\color{blue}We assume that the phase shifts can be continuously varied in $[0,2\pi)$ to characterize the fundamental performance of RIS. Although using the RIS with discrete phase shifts causes performance loss \cite{wu2020beamforming}, our results serve as performance upper bounds for the considered system which are still useful for characterizing the fundamental performance of the RIS-aided system.}}, ${\beta _\ell} \in (0,1]$ represents the amplitude reflection coefficient, ${\phi _\ell} = \exp \left( {j{\theta _\ell}} \right)$, $n=1,\cdots,L$, ${\theta _\ell} \in [0,2\pi ),$ denotes the phase shift. 
	
	{\color{blue}In the literature, numerous methods have been proposed to obtain the effective phase shift introduced by all the elements of the RIS, e.g., \cite{wu2019intelligent,yu2020optimal,elbir2020deep,du2020millimeter, hou2019mimo}. To mitigate the interference at the $n_{\rm th}$ user and perform well both in RIS and non-RIS scenarios, we adopt the design of passive beamforming process and the downlink user-detection vectors proposed in \cite{hou2019mimo}. The passive beamforming matrix at RIS can be obtained as \cite[eq. (13)]{hou2019mimo}
	\begin{equation}
	{\bf{\Phi }} =\frac{{\bf{\tilde H}}^{ - 1}{\bf{X}}}{\beta _{\max }},
	\end{equation}
	where $\beta _{\max }$ is obtained by finding the maximum amplitude coefficient, ${\bf{X}}$ is the signal sent by BS, and ${{{\bf{\tilde H}}}}$ can be formulated from ${{\bf{Q}}_{u_R-u_1,{\rm L}}}{\bf{\Phi }}{{\bf{Q}}_{u_R,{\rm L}}}$, $\cdots$, ${{\bf{Q}}_{u_R-u_N,{\rm L}}}{\bf{\Phi }}{{\bf{Q}}_{u_R,{\rm L}}}$ as \cite[eq. (9)]{hou2019mimo}.	
	
	The detection vector of the $n_{\rm th}$ user can be written as \cite[eq. (19)]{hou2019mimo}
	\begin{equation}
	{{\bf{v}}_{{u_n}}} = {{\bf{T}}_{{u_n}}}{{\bf{x}}_{{u_n}}},
	\end{equation}
	where ${{\bf{T}}_{{u_n}}}$ is derived by the left singular vectors of ${{{\bf{\tilde H}}}}$. With the aid of the classic maximal ratio combining technique, ${{\bf{x}}_{{u_n}}}$ can be expressed as
	\begin{equation}
	{{\bf{x}}_{{u_n}}} = \frac{{{\bf{T}}_{{u_n}}^{\bf{H}}{{\bf{h}}_{{u_n}}}}}{{\left| {{\bf{T}}_{{u_n}}^{\bf{H}}{{\bf{h}}_{{u_n}}}} \right|}},
	\end{equation}
	where ${{\bf{h}}_{{u_n}}}$ is the ${u_n}_{\rm th}$ column of the effective channel matrix ${{\bf{Q}}_{u_R-u_n,{\rm L}}}{\bf{\Phi }}{{\bf{Q}}_{u_R,{\rm L}}}$.
	Thus, the SINRs involved in both scenarios can be obtained.}
	
	For the first case, the received SINR at user $n$ and the eavesdropper are denoted as ${\rm SINR}_{\upsilon,\sigma}$, and it can be expressed as\footnote{\color{blue}{To facilitate effective eavesdropping, the eavesdropper should try to close to BS  because the SINR of the eavesdropper increases as $d_{u_e,\sigma}$ decreases.}}
	\begin{equation}\label{losusersinr}
	{{\rm SINR}_{\upsilon,\sigma}} = \frac{{{{{\cal G}_\upsilon }}\left\| {{{\widehat {\bf{h}}}_{\upsilon,\sigma}}} \right\|_F^2{{\left( {{d_{\upsilon,\sigma}}} \right)}^{ - \alpha_\sigma}}{p_{u_n}}}}{{\beta _{\max }^2{Q}^2{\sigma_N^{2}}}},
	\end{equation}
	{\color{blue}where $p_{u_n}$ denotes the transmit power of the BS for user, $Q \triangleq K-M-1$ is the effective antenna gain, and}
	\begin{equation}
	{{{\widehat {\bf{h}}}_{\upsilon,\sigma}}}= \left[ {\begin{array}{*{20}{c}}
	{{q_{1,1}^{\upsilon,\sigma}}}\\
	\vdots \\
	{{q_{Q,{1}}^{\upsilon,\sigma}}}
	\end{array}\begin{array}{*{20}{c}}
	\cdots \\
	\vdots \\
	\cdots
	\end{array}\begin{array}{*{20}{c}}
	{{q_{1,M}^{\upsilon,\sigma}}}\\
	\vdots \\
	{{q_{Q,M}^{\upsilon,\sigma}}}
	\end{array}} \right]
	\end{equation}
	is a $Q \times M$ matrix which denotes the channel gain of user $n$. In addition, we can see that the eavesdropper's SINR is affected by the random directivity gain ${G_{{u_e}}(\theta )}$.
	
	For the second case, the SINR of LoS path can be obtained by letting $\sigma={\rm L}$ in \eqref{losusersinr}. For the RIS-aided path, with the help of the design of passive beamforming process \cite{hou2019mimo}, the SINR with the optimal phase shift design of RIS’s reflector array can be derived as \cite[eq. (29)]{hou2019mimo}
	\begin{equation}\label{SINRlis}
	{{\rm SINR}_{\upsilon,{\rm RIS}}}\! = \!\frac{{{\mathcal{G}_{\upsilon}}C_{L_1}C_{L_2}\left\| {{{\widehat {\bf{h}}}_{\upsilon,{\rm RIS}}}} \right\|_F^2{{\left( {{d_{u_R,{\rm L}}}{d_{\upsilon,{\rm L}}}} \right)}^{ - 2}}{p_{u_n}}}}{{\beta _{\max }^2{Q}^2{\sigma_N^{2}}}},
	\end{equation}
	{\color{blue}where $\upsilon = u_R-u_n, u_R-u_e$}, and
	\begin{equation}
	{{{\widehat {\bf{h}}}_{\upsilon,{\rm RIS}}}}\!  = \!\!\left[\!\! {\begin{array}{*{20}{c}}
	{\left( {{q_{1,1}^{\upsilon,{\rm L}}}} \right) \cdot \left( {{q_{1,m}^{u_R,{\rm L}}}} \right)}& \cdots &{\left( {{q_{1,L}^{\upsilon,{\rm L}}}} \right) \cdot \left( {{q_{L,m}^{u_R,{\rm L}}}} \right)}\\
	\vdots & \vdots & \vdots \\
	{\left( {{q_{Q,1}^{\upsilon,{\rm L}}}} \right) \cdot \left( {{q_{1,m}^{u_R,{\rm L}}}} \right)}& \cdots &{\left( {{q_{Q,L}^{\upsilon,{\rm L}}}} \right) \cdot \left( {{q_{L,m}^{u_R,{\rm L}}}} \right)}
	\end{array}} \!\!\right]
	\end{equation}
	is a $Q\times L$ matrix.
	
	With the help of \eqref{losusersinr} and \eqref{SINRlis}, we derive the exact PDF and CDF expressions for each scenario's SINR in terms of multivariate Fox's $H$-function \cite[eq. (A-1)]{mathai2009h}, which are summarized in the following Theorems.	
	\begin{them}\label{P1}
	For the first communication scenario, let $Z_{\upsilon,\sigma}= {{\rm SINR}_{\upsilon,\sigma}} $, the CDF of $Z_{\upsilon,\sigma}$ is expressed as
	\begin{equation}\label{FinalSINR1}
	F_{Z_{\upsilon,\sigma}}(z) = \prod\limits_{\ell  = 1}^{QM} {\frac{1}{{\Gamma \left( {{m_{s_\ell}}} \right)\Gamma \left( {{m_\ell }} \right)}}} (H_{{{\rm CDF}}_1,2} -H_{{{{\rm CDF}}_1},0}) ,
	\end{equation}
	where $H_{{{\rm CDF}}_1,\hbar}$ $(\hbar=0,2)$ is derived as \eqref{H1} at the top of the next page, ${A_1}=\frac{{{\mathcal{G}_{\upsilon}}{p_{{u_n}}}}}{{{\beta _{\max}^2}{Q^2}\sigma_N^2}}$,
	$\left\{ {{m_\ell }} \right\}_{i = 1}^{QM} = \left\{ {\left\{ {{m_{q,n}^{\upsilon,\sigma }}} \right\}_{n = 1}^M} \right\}_{q = 1}^Q{\rm{ }}$, $
	\left\{ {{m_{{s_\ell }}}} \right\}_{i = 1}^{QM} \!=\! \left\{ {\left\{ {{m_s}_{q,n}^{\upsilon,\sigma }} \right\}_{n = 1}^M} \right\}_{q = 1}^Q{\rm{ }}$, and $\left\{ {{\Omega _\ell }} \right\}_{i = 1}^{QM} \!=\! \left\{ {\left\{ {{\Omega _{q,n}^{\upsilon,\sigma }}} \right\}_{n = 1}^M} \right\}_{q = 1}^Q.$
	\newcounter{mytempeqncnt}
	\begin{figure*}[t]
	\normalsize
	\setcounter{mytempeqncnt}{\value{equation}}
	\setcounter{equation}{22}
	{\small \begin{align}\label{H1}
	H_{{{\rm CDF}}_1,\hbar} \triangleq \frac{{2r_\hbar ^2}}{{\left( {r_2^2 - r_0^2} \right)\alpha_\sigma }} H_{{1},{2}:2,1; \cdots ;2,1}^{0,{1}:1,2; \cdots ;1,2}\!\!\left(\! {\left. \!\!\!{\begin{array}{*{20}{c}}
	{\frac{{{A_1}zr_\hbar ^{\alpha_\sigma} {m_1}}}{{{{\bar \gamma }_1}\left( {{m_{{s_1}}} - 1} \right)}}}\\
	\vdots \\
	{\frac{{{A_1}zr_\hbar ^{\alpha_\sigma} {m_{QM}}}}{{{{\bar \gamma }_{QM}}\left( {{m_{{s_{QM}}}} - 1} \right)}}}
	\end{array}} \!\!\right|\!\!\!\begin{array}{*{20}{c}}
	{\left( \!{1 \!- \!\frac{2}{{\alpha_\sigma} };1, \!\cdots \!,1} \!\right):\left( {1,1} \right),\left( {1 \!- \!{m_{{s_1}}},1} \!\right); \!\cdots\! ;\left(\!{1,1} \!\right),\left(\! {1 \!- \!{m_{{s_{QM}}}},1} \right)}\\
	{\left(\! {0;1, \!\cdots\! ,1} \!\right)\left( { - \!\frac{2}{{\alpha_\sigma} };1, \cdots ,1}\! \right):\left( \!{{m_1},1}\! \right); \!\cdots \!;\left( {{m_{QM}},1} \!\right)}
	\end{array}}\! \!\!\!\right).
	\end{align}}
	\setcounter{equation}{\value{mytempeqncnt}}
	\hrulefill
	\end{figure*}
	\setcounter{equation}{23}
	
	\begin{IEEEproof}
	Please refer to Appendix \ref{AppendixA}.
	\end{IEEEproof}
	\end{them}
	\begin{them}\label{P3}
	For the first communication scenario, the PDF of the SINR for the user $n$ and eavesdropper can be given by
	\begin{equation}\label{FinalSINR1PDF}
	f_{Z_{\upsilon,\sigma}}(z)=\prod\limits_{\ell  = 1}^{QM} {\frac{1}{{\Gamma \left( {{m_{s_\ell}}} \right)\Gamma \left( {{m_\ell }} \right)}}} (H_{{{\rm PDF}}_1,2} -H_{{{\rm PDF}}_1,0}),
	\end{equation}
	where $H_{{{\rm PDF}}_1,\hbar}$ $(\hbar=0,2)$ is derived as \eqref{H2} at the top of the next page.
	\newcounter{mytempeqncnt4}
	\begin{figure*}[t]
	\normalsize
	\setcounter{mytempeqncnt4}{\value{equation}}
	\setcounter{equation}{24}
	{\small \begin{align}\label{H2}
	H_{{{\rm PDF}}_1,\hbar} \!\triangleq \!\frac{{2r_\hbar ^{2{+}\alpha }}{A_1}}{{\left( {r_2^2 - r_0^2} \right)\alpha }} H_{{1},{2}:2,1; \cdots ;2,1}^{0,{1}:1,2; \cdots ;1,2}\left(\!\!\!\! {\left. {\begin{array}{*{20}{c}}
	{\frac{{{A_1}zr_\hbar ^\alpha {m_1}}}{{{{\bar \gamma }_1}\left( {{m_{{s_1}}} - 1} \!\right)}}}\\
	\vdots \\
	{\frac{{{A_1}zr_\hbar ^\alpha {m_{QM}}}}{{{{\bar \gamma }_{QM}}\left( {{m_{{s_{QM}}}} - 1} \!\right)}}}
	\end{array}} \!\!\!\!\right|\!\!\!\!\begin{array}{*{20}{c}}
	{\left( { - \frac{2}{\alpha };1, \!\cdots \!,1}\! \right):\left( {1,1} \right),\left(\! {1\! -\! {m_{{s_1}}},1} \!\right);\! \cdots \!;\left( {1,1} \right),\left(\! {1 \!- \!{m_{{s_{QM}}}},1}\! \right)}\\
	{\left( {{1};1,\! \cdots\! ,1} \right)\left( { - 1 \!-\! \frac{2}{\alpha };1,\! \cdots\! ,1} \right):\left( {{m_1},1} \right); \!\cdots\! ;\left( {{m_{QM}},1} \right)}
	\end{array}}\!\!\!\! \right).
	\end{align}}
	\setcounter{equation}{\value{mytempeqncnt4}}
	\hrulefill
	\end{figure*}
	\setcounter{equation}{25}
	\begin{IEEEproof}
	Following similar procedures as in Appendix \ref{AppendixB}, we can derive the PDF of $Z_{\upsilon,\sigma}$ by calculating
	\begin{equation}\label{PDFjisuna}
	{f_Y}(y) = \int_{r_0^{{\alpha _\sigma }}}^{r_2^{{\alpha _\sigma }}} x {f_X}(xy){f_D}(x){\rm d}x,
	\end{equation}
	where ${f_X}(\cdot)$ is given as \cite[eq. (23)]{rahama2018sum} and ${f_D}(\cdot)$ has been derived as \eqref{ddeaerfacifang}.
	\end{IEEEproof}	
	\end{them}
	Then, we focus on the derivation for the performance metric in the second scenario.	
	\begin{them}\label{P2}
	For the second communication scenario, let $Z_{\upsilon,{\rm RIS}}= {{\rm SINR}_{\upsilon,{\rm RIS}}} $, the CDF of $Z_{\upsilon,{\rm RIS}}$ can be expressed as
	\begin{equation}\label{RISdeCDF}
	F_{Z_{\upsilon,{\rm RIS}}}(z) =\prod\limits_{\ell  = 1}^{QL} {\frac{H_{{{\rm CDF}}_2,2}-H_{{{\rm CDF}}_2,0}}{{\prod\limits_{i = 1}^2 {\Gamma \left( {{m_{{s_{i,\ell }}}}} \right)\Gamma \left( {{m_{i,\ell }}} \right)} }}} ,
	\end{equation}
	where $H_{{{\rm CDF}}_2,\hbar} $ $(\hbar=0,2)$ is derived as \eqref{H3} at the top of the next page, $\Delta \! =\! {\left(\! {\frac{{{A_2}zr_2^2 {m_{1,1}}{m_{2,1}}}}{{\prod\limits_{i = 1}^2 {\left( {{m_{{s_{i,1}}}} - 1} \right)} {{\bar \gamma }_1}}}, \cdots ,\frac{{{A_2}zr_2^2{m_{1,QL}}{m_{2,QL}}}}{{\prod\limits_{i = 1}^2 {\left( {{m_{{s_{i,QL}}}} - 1} \right)} {{\bar \gamma }_{QL}}}}} \!\right)^{\bf T}}$,
	${A_2}=\frac{{{\mathcal{G}_{\upsilon}}{C_{L_1} }{C_{L_2} }(d_{u_R,{\rm L}})^{-2}{p_{{u_n}}}}}{{{\beta _{\max}^2}{Q^2}\sigma_N^2}}$, $\left\{\! {{m_{1,\ell }}} \!\right\}_{i = 1}^{QL} \!= \!\left\{ {\left\{\! {{m_{q,n}^{\upsilon,{\rm L}}}}\! \right\}_{n = 1}^M} \right\}_{q = 1}^Q{\rm{ }}$, $\left\{\! {{m_{1,{s_\ell }}}} \!\right\}_{i = 1}^{QL} = \!\left\{ \!{\left\{\! {{m_s}_{q,n}^{\upsilon,{\rm L}}}\! \right\}_{n = 1}^M} \!\right\}_{q = 1}^Q{\rm{ }}$, $\left\{ \!{{\Omega _{1,\ell }}}\! \right\}_{i = 1}^{QL} = \left\{\! {\left\{ {{\Omega _{q,n}^{\upsilon,{\rm L}}}} \right\}_{n = 1}^M} \!\right\}_{q = 1}^Q$, $\left\{\! {{m_{2,\ell }}} \!\right\}_{i = 1}^{L} = \left\{\! {{m_{n,m}^{u_R,{\rm L}}}} \!\right\}_{n = 1}^{L}$, $\left\{ {{m_{2,{s_\ell }}}} \right\}_{i = 1}^{L} = \left\{ {{m_s}_{n,m}^{u_R,{\rm L}}} \right\}_{n = 1}^{L}$, and $\left\{ {{\Omega _{2,\ell }}} \right\}_{i = 1}^{L} = \left\{ {{\Omega _{n,m}^{u_R,{\rm L}}}} \right\}_{n = 1}^{L}.$
	\newcounter{mytempeqncnt5}
	\begin{figure*}[t]
	\normalsize
	\setcounter{mytempeqncnt5}{\value{equation}}
	\setcounter{equation}{27}
	{\small \begin{align}\label{H3}
	H_{{{\rm CDF}}_2,\hbar} \triangleq \frac{{r_\hbar ^{2 }}}{{\left( {r_2^2 - r_0^2} \right) }} H_{{1},{2}:2,1; \cdots ;2,1}^{0,{1}:1,2; \cdots ;1,2}\!\!\left(\!\! {\Delta \!\left|\!\!\! {\begin{array}{*{20}{c}}
	{\left(\! {0 ;1,\! \cdots\! ,1} \right)\!:\!\left( {1,1} \!\right),\left\{ {\left(\! {1 -\! {m_{{s_{i,1}}}},1} \!\right)} \right\}_{i = 1}^2; \!\cdots\! ;\left(\! {1,1} \!\right),\left\{ {\left(\! {1 - {m_{{s_{i,QL}}}},1}\! \right)} \right\}_{i = 1}^2}\\
	{\left(\! {0;1, \!\cdots \!,1}\! \right)\left(\! { - \frac{2}{2 };1, \!\cdots ,1} \right)\!:\!\left\{ {\left( {{m_{i,1}},1} \right)} \right\}_{i = 1}^2;\! \cdots \!;\left\{ {\left(\! {{m_{i,QL}},1} \!\right)} \right\}_{i = 1}^2}
	\end{array}} \right.}\!\!\!\right).
	\end{align}}
	\setcounter{equation}{\value{mytempeqncnt5}}
	\hrulefill
	\end{figure*}
	\setcounter{equation}{28}
	\begin{IEEEproof}
	Please refer to Appendix \ref{AppendixB}.
	\end{IEEEproof}
	\end{them}
	\begin{them}\label{P4}
	For the second communication scenario, the PDF of $Z_{\upsilon,{\rm RIS}}$ can be deduced as
	\begin{equation}
	f_{Z_{\upsilon,{\rm RIS}}}(z) =\prod\limits_{\ell  = 1}^{QL} {\frac{(H_{{{\rm PDF}}_2,2}-H_{{{\rm PDF}}_2,0})}{{\prod\limits_{i = 1}^2 {\Gamma \left( {{m_{{s_{i,\ell }}}}} \right)\Gamma \left( {{m_{i,\ell }}} \right)} }}} ,
	\end{equation}
	where $H_{{{\rm PDF}}_2,\hbar}$ $(\hbar=0,2)$ is derived as \eqref{H4} at the top of the next page.
	\newcounter{mytempeqncnt6}
	\begin{figure*}[t]
	\normalsize
	\setcounter{mytempeqncnt6}{\value{equation}}
	\setcounter{equation}{29}
	{\small \begin{align}\label{H4}
	H_{{{\rm PDF}}_2,\hbar} \triangleq \frac{{r_\hbar ^{4}}A_2}{{\left( {r_2^2 - r_0^2} \right) }} H_{{1},{2}:2,1; \cdots ;2,1}^{0,{1}:1,2;\!\cdots\! ;1,2}\left( \!\!{\Delta \!\left|\!\! \!{\begin{array}{*{20}{c}}
	{\left( { - 1;1,\!\cdots\! ,1} \right)\!:\!\left( {1,1} \right),\left\{ {\left( {1 - {m_{{s_{i,1}}}},1} \right)} \right\}_{i = 1}^2;\! \cdots\! ;\left( {1,1} \right),\left\{ {\left( \!{1 - {m_{{s_{i,QL}}}},1}\! \right)} \right\}_{i = 1}^2}\\
	{\left(\! {1;1, \!\cdots\! ,1}\! \right)\left( { - 2;1, \!\cdots\! ,1} \right)\!:\!\left\{ {\left(\! {{m_{i,1}},1} \right)} \!\right\}_{i = 1}^2; \!\cdots\! ;\left\{ {\left( {{m_{i,QL}},1} \right)} \right\}_{i = 1}^2}
	\end{array}} \right.}\!\!\!\! \right).
	\end{align}}
	\setcounter{equation}{\value{mytempeqncnt6}}
	\hrulefill
	\end{figure*}
	\setcounter{equation}{30}
	\begin{IEEEproof}
	Following similar procedures as in Appendix \ref{AppendixB}, we can express the PDF of $Z_{\upsilon,{\rm RIS}}$ with the help of \eqref{PDFjisuna}.
	\end{IEEEproof}	
	\end{them}
	\begin{rem}\label{RISshi}
	By comparing Theorem \ref{P1} and Theorem \ref{P2} or Theorem \ref{P3} and Theorem \ref{P4}, we can see that, caused by the characteristics of Fisher-Snedecor $\mathcal{F}$ distribution, the CDF and PDF of the SINR for the RIS-aided path are very similar to that of the LoS or NLoS path. This insight is very useful because we only need to further investigate Theorem \ref{P1} and Theorem \ref{P3} for performance analysis of the first communication scenario. In other words, we can get the results for the RIS-aided scenario through simple parameter transformation.
	Specifically, we can let $m_\ell \to m_{1,\ell} m_{2,\ell}$, $m_{s_\ell} \to m_{s_{1,\ell}} m_{s_{2,\ell}}$, $\Gamma \left( {m_\ell} \right) \to \Gamma \left( m_{1,\ell} \right) \Gamma \left( m_{2,\ell} \right)$, $\Gamma \left( {m_{s_\ell}} \right) \to \Gamma \left( m_{s_{1,\ell}} \right) \Gamma \left( m_{s_{2,\ell}} \right)$, $(m_\ell) \to (m_{1,\ell});(m_{2,\ell})$, $(m_{s_\ell}) \to (m_{s_{1,\ell}});(m_{s_{2,\ell}})$, $QM \to QL$ and $\alpha_\sigma={\rm NL}$, where $A \to B$ means replacing $A$ with $B$, and the results for the RIS-aided scenario will follow.
	\end{rem}
	
	\section{Security Performance Analysis}\label{sec3}
	The secrecy rate over fading wiretap channels \cite{bloch2008wireless} is defined as the difference between the main channel rate and the wiretap channel rate as
	\begin{align}
	&{C_s}\left( {{Z_{{u_n},\sigma} },{Z_{{u_e},\sigma} }} \right) = \left\{ {\begin{array}{*{20}{l}}
	{{C_{{u_n},\sigma}} - {C_2},}&{{{Z_{{u_n},\sigma} }} > {Z_{{u_e},\sigma}.}}\\
	{0,}&{{\rm{ otherwise,}}}
	\end{array}} \right. \notag\\
	&= \left\{ {\begin{array}{*{20}{l}}
	{{{\log }_2}\left( {\frac{{1 + {{Z_{{u_n},\sigma} }}}}{{1 + {{Z_{{u_e},\sigma}}}}}} \right),}&{{{Z_{{u_n},\sigma} }} > {Z_{{u_e},\sigma}}.}\\
	{0,}&{{\rm{ otherwise,}}}
	\end{array}} \right.
	\end{align}
	which means that a positive secrecy rate can be assured if and only if the received SINR at user $n$ has a superior quality than that at the eavesdropper.
	
	In the considered RIS-aided system, we assume that the location of users are random and define
	\begin{equation}
	{\bf{P_A}} \triangleq \left[ {{P_{\rm LoS}},{P_{\rm NLoS}}} \right]
	\end{equation}
	and
	\begin{equation}
	{\bf{P_B}}\! \!\triangleq \!\left[ {{{\Pr}_{\rm LoS}}{{\Pr}_G},\!{{\Pr}_{\rm LoS}}{{\Pr}_g},\!{{\Pr}_{\rm NLoS}}{{\Pr}_G},\!{{\Pr}_{\rm NLoS}}{{\Pr}_g}} \right],
	\end{equation}
	where ${{\Pr}_{\rm LoS}} \triangleq {B_1}{B_2}$, ${{\Pr}_{\rm NLoS}} \triangleq 1 - {B_1}{B_2}$, ${{\Pr}_G}\triangleq \frac{{{\theta _c}}}{{180}}$, ${{\Pr}_g} \triangleq 1 - \frac{{{\theta _c}}}{{180}}$, $P_{\rm LoS}$ represents the probability that the path is LoS in two scenarios, $P_{\rm NLoS}$ denotes the probability that the path is NLoS in the first scenario or RIS-aided in the second scenario, ${P_G}$ is the probability that the eavesdropper's directivity gain is the same as the user's, and ${P_g}$ represents the probability that the eavesdropper's directional gain is ${\bf g}_{u_e}$. Besides, we assume that the eavesdropper has the same path loss as user $n$ because they are close to each other.
	\subsection{Outage Probability Characterization}\label{S1OP}
	{\textcolor{blue}{The OP is defined as the probability that the instantaneous SINR is less than $Z_{\rm th}$, where $Z_{\rm th}$ is the determined SNR threshold.}} In the first communication scenario, the OP can be directly calculated as
	\begin{equation}\label{OP}
	OP={\bf {P_A}} {\left[ {{F_{{Z_{{u_n},{\rm L}}}}}(Z_{\rm th}),{F_{{Z_{{u_n},{\rm NL}}}}}(Z_{\rm th})} \right]^{\rm  T}},
	\end{equation}
	which can be evaluated directly with the help of \eqref{FinalSINR1}.
	\begin{rem}
	In the second communication scenario, the OP can be obtained by replacing ${F_{{Z_{{u_n},{\rm NL}}}}}(Z_{\rm th})$ in \eqref{OP} with ${F_{{Z_{{u_n},{\rm RIS}}}}}(Z_{\rm th})$. We can observe that OP decreases when the channel condition of user $n$ is improved. Moreover, a RIS equipped with more elements will also make the OP lower.
	\end{rem}
	\subsection{SOP Characterization}\label{S2SOP}
	The secrecy outage probability (SOP), is defined as the probability that the instantaneous secrecy capacity falls below a target secrecy rate threshold.
	In the first communication scenario, the SOP can be written as
	\begin{align}\label{SOP}
	SOP &= {\bf{P_B}}\left[ \begin{array}{l}
	\Pr \left( {{Z_{{u_n},{\rm L}}} \le {R_s}{Z_{{u_e},{\rm L},G}} + {R_s} - 1} \right)\\
	\Pr \left( {{Z_{{u_n},{\rm L}}} \le {R_s}{Z_{{u_e},{\rm L},g}} + {R_s} - 1} \right)\\
	\Pr \left( {{Z_{{u_n},{\rm NL}}} \le {R_s}{Z_{{u_e},{\rm NL},G}} + {R_s} - 1} \right)\\
	\Pr \left( {{Z_{{u_n},{\rm NL}}} \le {R_s}{Z_{{u_e},{\rm NL},g}} + {R_s} - 1} \right)
	\end{array} \right]\notag\\
	&={\bf{P_B}}{\left[ {SO{P_{1,G}},SO{P_{1,g}},SO{P_{2,G}},SO{P_{2,G}}} \right]^{\rm T}},
	\end{align}
	where {\color{blue}$R_{t}$ is the target secrecy rate, $R_{s}=2^{R_{t}}$ \cite{kong2018physical}}, ${Z_{{u_e},{\rm L},G}}$ denotes that the eavesdropper's directional gain is  ${G}_{u_e}$ and ${Z_{{u_e},{\rm L},g}}$ denotes that the eavesdropper's directional gain is  ${g}_{u_e}$.
	With the help of \eqref{SOP} and Theorems \ref{P1}-\ref{P4}, we derive the following propositions.
	\begin{prop}\label{PC}
	Let $SOP_{\sigma,{\mathchar'26\mkern-10mu\lambda} } \!\triangleq\! \Pr \left(\! {{Z_{{u_n},\sigma}} \!\le \! {R_s}{Z_{{u_e},\sigma,{\mathchar'26\mkern-10mu\lambda}}} \!+\! {R_s} \!- \!1} \!\right)$ denotes the element of the matrix in \eqref{SOP}, where $\mathchar'26\mkern-10mu\lambda=G, g$. We can express $SOP_{\sigma,{\mathchar'26\mkern-10mu\lambda} }$ as
	\begin{equation}\label{SIOsolve}
	SOP_{\sigma,{\mathchar'26\mkern-10mu\lambda} }={{\cal S}_{2,2}} - {{\cal S}_{2,0}} - {{\cal S}_{0,2}} + {{\cal S}_{0,0}},
	\end{equation}
	where ${{\cal S}_{p,q}}$ $(p=0,2$ and $q=0,2)$ is derived as \eqref{sdaff} at the top of the next page, ${H_{SOP}} \triangleq{\left( {\frac{{{A_1}zr_p^\alpha {m_1}}}{{{{\bar \gamma }_1}\left( {{m_{{s_1}}} - 1} \right)}}, \cdots ,\frac{{{A_1}zr_q^\alpha {m_{2QM + 1}}}}{{{{\bar \gamma }_{2QM + 1}}\left( {{m_{{s_{2QM + 1}}}} - 1} \right)}},e} \right)^T}$, ${H_{{S_1}}}$ and ${H_{{S_1}}}$ are respectively given as \eqref{akfhas} and \eqref{akfhas2} at the top of this page, and $e$ is a positive number close to zero (e.g., $e = 10^{-6}$).
	\newcounter{mytempeqncnt1}
	\begin{figure*}[t]
	\normalsize
	\setcounter{mytempeqncnt1}{\value{equation}}
	\setcounter{equation}{36}
	{\small \begin{align}\label{sdaff}
	{{\cal S}_{p,q}} = \frac{{4{A_1}{r_p}^2r_q^{2 + {\alpha _\sigma }}}}{{{\alpha _{{u_n}}}{\alpha _{{u_e}}}{{\left( {r_2^2 - r_0^2} \right)}^{2}}}}\times H_{4,4:2,1; \cdots ;2,1;0,1}^{0,3:1,2; \cdots ;1,2;1,0}\left( {{H_{SOP}}\left| {\begin{array}{*{20}{c}}
	{{H_{{S_1}}}:\left( {1,1} \right),\left( {1 - {m_{{s_1}}},1} \right); \cdots ;\left( {1,1} \right),\left( {1 - {m_{{s_{2QM}}}},1} \right); - }\\
	{{H_{{S_2}}}:\left( {{m_1},1} \right); \cdots ;\left( {{m_{2QM}},1} \right);\left( {0,1} \right)}
	\end{array}} \right.} \right).
	\end{align}}
	{\small \begin{align}\label{akfhas}
	{H_{{S_1}}} \triangleq\left( {1 - \frac{2}{\alpha };\underbrace {1, \cdots ,1}_{QM},\underbrace {0, \cdots ,0}_{QM},0} \right);\left( { - \frac{2}{\alpha };\underbrace {0, \cdots ,0}_{QM},\underbrace {1, \cdots ,1}_{QM},0} \right);\left( {2,\underbrace { - 1, \cdots , - 1}_{{2}QM + 1}} \right);\left( {0,\underbrace {1, \!\cdots\! ,1}_{{2}QM + 1}} \right).
	\end{align}}
	{\small \begin{align}\label{akfhas2}
	{H_{{S_2}}} \!\triangleq \!\left(\! {0;\underbrace {1,\! \cdots\! ,1}_{QM},\underbrace {0, \!\cdots\! ,0}_{QM},0} \!\right)\left(\! { - \frac{2}{\alpha };\underbrace {1,\! \cdots\! ,1}_{QM},\underbrace {0, \!\cdots\! ,0}_{QM},0} \!\right)\!;\!\left(\! {1;\underbrace {0, \cdots ,0}_{QM};\underbrace {1, \!\cdots\! ,1}_{QM},0}\! \right)\!;\!\left( \!{ - 1 - \frac{2}{\alpha };\underbrace {0, \!\cdots\! ,0}_{QM},\underbrace {1,\! \cdots \!,1}_{QM},0} \!\right).
	\end{align}}
	\setcounter{equation}{\value{mytempeqncnt1}}
	\hrulefill
	\end{figure*}
	\setcounter{equation}{39}
	\begin{IEEEproof}
	Please refer to Appendix \ref{AppendixC}.
	\end{IEEEproof}
	\begin{rem}
	In the second communication scenario, the SOP of the RIS-aided path can be obtained with the help of Remark \ref{RISshi}. From \eqref{SIOsolve}, we can see that the SOP will decrease when the fading parameter $m_{u_n}$ and $m_{s_{u_n}}$ increase because of better communication conditions. In addition, we can see that $Q$, $M$, and $L$ will affect the dimension of the multivariate Fox's $H$-function, so $Q$, $M$ and $L$ have a greater impact on the SOP than the channel parameters. Thus, it is obvious that communication system designers can increase $L$ of RIS for lower SOP.
	\end{rem}
	\end{prop}
	\subsection{PNSC Characterization}\label{S3PNSC}
	Another fundamental performance metric of the PLS is the probability of non-zero secrecy capacity (PNSC) which can be defined as
	\begin{align}\label{PNSCCHUHSI}
	&PNSC = {\bf{P_B}}\left[ \begin{array}{l}
	\Pr \left( {{Z_{{u_n},{\rm L}}} > {Z_{{u_e},{\rm L},G}}} \right)\\
	\Pr \left( {{Z_{{u_n},{\rm L}}} > {Z_{{u_e},{\rm L},g}}} \right)\\
	\Pr \left( {{Z_{{u_n},{\rm NL}}} > {Z_{{u_e},{\rm NL},G}}} \right)\\
	\Pr \left( {{Z_{{u_n},{\rm NL}}} > {Z_{{u_e},{\rm NL},g}}} \right)
	\end{array} \right]\notag\\
	&={\bf{P_B}}{\left[ {PN{Z_{1,G}},PN{Z_{1,g}},PN{Z_{2,G}},PN{Z_{2,G}}} \right]^{\rm T}}.
	\end{align}
	Thus, we exploit \eqref{PNSCCHUHSI} and Theorems \ref{P1}-\ref{P4} to arrive the following proportions.
	\begin{prop}\label{PD}
	Let $PNSC_{\sigma,{\mathchar'26\mkern-10mu\lambda} } \triangleq \Pr \left( {{Z_{{u_n},\sigma}} > {R_s}{Z_{{u_e},\sigma,{\mathchar'26\mkern-10mu\lambda}}}} \right)$ denotes the element of the matrix in \eqref{PNSCCHUHSI} and we can express $PNSC_{\sigma,{\mathchar'26\mkern-10mu\lambda} }$ as
	\begin{equation}\label{PNSCsolve}
	PNSC_{\sigma,{\mathchar'26\mkern-10mu\lambda} }={{\cal P}_{2,2}} - {{\cal P}_{2,0}} - {{\cal P}_{0,2}} + {{\cal P}_{0,0}},
	\end{equation}
	where ${{\cal P}_{p,q}}$ $(p=0,2$ and $q=0,2)$ can be written as \eqref{kokao} at the top of the next page, ${H_{PNSC}}\!\triangleq\!{\left(\! {\frac{{{A_1}zr_p^\alpha {m_1}}}{{e{{\bar \gamma }_1}\left(\! {{m_{{s_1}}} \!- \!1}\! \right)}}, \!\cdots \!,\frac{{{A_1}zr_q^\alpha {m_{2QM}}}}{{e{{\bar \gamma }_{2QM}}\left( \!{{m_{{s_{2QM}}}}\! - \!1} \!\right)}}} \right)^{\rm T}}$, ${H_{{P_1}}}$ and ${H_{{P_2}}}$ are expressed as \eqref{avmelmvla} and \eqref{avmelmvla2} at the top of the next page, respectively.
	\newcounter{mytempeqncnt2}
	\begin{figure*}[t]
	\normalsize
	\setcounter{mytempeqncnt2}{\value{equation}}
	\setcounter{equation}{41}
	{\small \begin{align}\label{kokao}
	{{\cal P}_{p,q}}&  = \frac{{4{A_1}{r_p}^2r_q^{2 + {\alpha _\sigma }}}}{{e{\alpha _{{u_n}}}{\alpha _{{u_e}}}{{\left( {r_2^2 - r_0^2} \right)}^{2}}}} H_{3,4:2,1; \cdots ;2,1}^{0,3:1,2; \cdots ;1,2}\left( {{H_{PNSC}}\left| {\begin{array}{*{20}{c}}
	{{H_{{S_1}}}:\left( {1,1} \right),\left( {1 - {m_{{s_1}}},1} \right); \cdots ;\left( {1,1} \right),\left( {1 - {m_{{s_{2QM}}}},1} \right)}\\
	{{H_{{S_2}}}:\left( {{m_1},1} \right); \cdots ;\left( {{m_{2QM}},1} \right)}
	\end{array}} \right.} \right).
	\end{align}}
	{\small \begin{align}\label{avmelmvla}
	{H_{{P_1}}} =\left( {1 - \frac{2}{\alpha }; \underbrace {1, \cdots,1}_{QM},\underbrace {0, \cdots ,0}_{QM}} \right);\left( { - \frac{2}{\alpha };\underbrace {0, \cdots ,0}_{QM},\underbrace {1,\cdots ,1}_{QM}} \right);\left( {0,\underbrace {1, \cdots ,1}_{{2}QM}} \right).
	\end{align}}
	{\small \begin{align}\label{avmelmvla2}
	{H_{{P_2}}} \triangleq \left( {0;\underbrace {1, \cdots ,1}_{QM},\underbrace {0, \cdots ,0}_{QM}}\! \right)\left(\! { - \frac{2}{\alpha };\underbrace {1, \cdots ,1}_{QM},\underbrace {0, \cdots ,0}_{QM}} \!\right);\left(\! {1;\underbrace {0, \cdots ,0}_{QM};\underbrace {1, \cdots ,1}_{QM}} \right)\!;\!\left(\! { - \!1 \!-\! \frac{2}{\alpha };\underbrace {0, \cdots ,0}_{QM},\underbrace {1, \cdots ,1}_{QM}} \!\right).
	\end{align}}
	\setcounter{equation}{\value{mytempeqncnt2}}
	\hrulefill
	\end{figure*}
	\setcounter{equation}{44}
	\begin{IEEEproof}	
	Please refer to Appendix \ref{AppendixD}.
	\end{IEEEproof}
	\begin{rem}
	For the second communication scenario, the PNSC of the RIS-aided path can be easily obtained according to Remark \ref{RISshi}. From \eqref{PNSCsolve}, we can see that PNSC will increase when user $n$ has good communication conditions or when the eavesdropper is in a bad communication environment. Besides, we can observe that PNSC is also more susceptible to $Q$, $M$, and $L$, because these parameters directly determine the dimension of the multivariate Fox's $H$-function. In general, system designers can improve PNSC by increasing the size of the RIS.
	\end{rem}
	\end{prop}
	\subsection{ASR Characterization}\label{S4ASR}
	The ASR describes the difference between the rate of the main channel and wiretap channel over instantaneous SINR can be expressed as
	\begin{align}\label{ASRCHUSHI}
	ASR &={\bf{P_B}}\left[ \begin{array}{l}
	E\left[ {{C_s}\left( {{Z_{{u_n},{\rm L}}},{Z_{{u_e},{\rm L},G}}} \right)} \right]\\
	E\left[ {{C_s}\left( {{Z_{{u_n},{\rm L}}},{Z_{{u_e},{\rm L},g}}} \right)} \right]\\
	E\left[ {{C_s}\left( {{Z_{{u_n},{\rm NL}}},{Z_{{u_e},{\rm NL},G}}} \right)} \right]\\
	E\left[ {{C_s}\left( {{Z_{{u_n},{\rm NL}}},{Z_{{u_e},{\rm NL},g}}} \right)} \right]
	\end{array} \right]\notag\\
	&={\bf{P_B}}{\left[ {AS{C_{1,G}},AS{C_{1,g}},AS{C_{2,G}},AS{C_{2,G}}} \right]^{\rm T}}.
	\end{align}
	We derive the following proposition using \eqref{ASRCHUSHI} and Theorems \ref{P1}-\ref{P4}.
	\begin{prop}\label{PE}
	We assume that $ASR_{\sigma,{\mathchar'26\mkern-10mu\lambda} }\!\!=\!\!E\!\left[ {{C_s}\!\left( {{Z_{{u_n},\sigma}}\!,\!{Z_{{u_e},\!\sigma,{\mathchar'26\mkern-10mu\lambda}}}} \right)} \!\right]$ is the element of the matrix in \eqref{ASRCHUSHI} and we can expressed $ASR_{\sigma,{\mathchar'26\mkern-10mu\lambda} }$ as
	\begin{align}\label{ASRzuizho}
	ASR_{\sigma,{\mathchar'26\mkern-10mu\lambda} }=&\sum\limits_{i = 1}^2 {\left( {{{\cal A}_{2,2,i}} - {{\cal A}_{2,0,i}} - {{\cal A}_{0,2,i}} + {{\cal A}_{0,0,i}}} \right)}\notag\\
	&  - \left( {{{\cal A}_{2,2,3}} - {{\cal A}_{2,0,3}} - {{\cal A}_{0,2,3}} + {{\cal A}_{0,0,3}}} \right),
	\end{align}
	where ${{\cal A}_{p,q,i}}$ $(p=0,2$, $q=0,2$, and $i=1,2)$ can be derived as \eqref{faejlak} at the top of the next page, ${H_{ASR}} \triangleq {\left( {\frac{{{A_1}zr_p^\alpha {m_1}}}{{{{\bar \gamma }_1}\left( {{m_{{s_1}}} - 1} \right)}}, \cdots ,\frac{{{A_1}zr_q^\alpha {m_{2QM + 1}}}}{{{{\bar \gamma }_{2QM + 1}}\left( {{m_{{s_{2QM + 1}}}} - 1} \right)}},e} \right)^T}$, ${H_{{A_1}}}$, ${H_{{A_2}}}$, and ${{\cal A}_{p,q,3}}$ $(p=0,2$ and $q=0,2)$ are respectively given as \eqref{ajfokwp2}, \eqref{ajfokwp3} and \eqref{aegae1} at the top of the next page, ${H_{AS{C_2}}}\! \triangleq\! {\left(\! {\frac{{{A_1}zr_p^\alpha {m_1}}}{{{{\bar \gamma }_1}\left( {{m_{{s_1}}} \!-\! 1} \right)}}, \!\cdots\! ,\frac{{{A_1}zr_q^\alpha {m_{QM + 1}}}}{{{{\bar \gamma }_{QM + 1}}\left( \!{{m_{{s_{QM + 1}}}} \!- \!1} \!\right)}},e} \!\right)^{\rm T}}$, ${H_{{A_{1,2}}}} \!\triangleq \!\left(\! { - \!\frac{2}{\alpha };\underbrace {0,\! \cdots\! ,0}_{QM},0}\! \right);\left(\! {1,\underbrace {1, \cdots ,1}_{QM},0} \!\right);\left(\! {0,\underbrace {1,\! \cdots\! ,1}_{QM},0} \!\right)$, and ${H_{{A_{2,2}}}}$ can be expressed as \eqref{afafarb}.
	\newcounter{mytempeqncnt3}
	\begin{figure*}[t]
	\normalsize
	\setcounter{mytempeqncnt3}{\value{equation}}
	\setcounter{equation}{46}
	{\begin{align}\label{faejlak}
	{{\cal A}_{p,q,i}}\! = \!\frac{{4{A_1}{r_p}^2r_q^{2 + {\alpha _\sigma }}}}{{{{e}\ln 2}{\alpha _{{u_n}}}{\alpha _{{u_e}}}{{\left( {r_2^2 - r_0^2} \right)}^{2}}}}H_{5,4:2,1; \cdots ;2,1;0,1}^{0,4:1,2; \cdots ;1,2;1,0}\left(\!\! {{H_{ASR}}\left| \!\!\!{\begin{array}{*{20}{c}}
	{{H_{{A_1}}}:\left( {1,1} \right),\left( {1 - {m_{{s_1}}},1} \right);\! \cdots \!;\left( {1,1} \right),\left( {1 - {m_{{s_{2QM}}}},1} \right); - }\\
	{{H_{{A_2}}}:\left( {{m_1},1} \right); \!\cdots\! ;\left( {{m_{2QM}},1} \right);\left( {1,1} \right)}
	\end{array}} \right.} \!\!\!\right).
	\end{align}}
	{\small \begin{align}\label{ajfokwp2}
	{H_{{A_1}}} \triangleq \left( {1 - \frac{2}{\alpha };\underbrace {1, \cdots ,1}_{QM},\underbrace {0, \cdots ,0}_{QM},0} \right);\left( { - \frac{2}{\alpha };\underbrace {0,\! \cdots \!,0}_{QM},\underbrace {1, \!\cdots\! ,1}_{QM},0} \right);\left( {1,\underbrace { - 1, \!\cdots\! , - 1}_{QM},\underbrace {1, \!\cdots \!,1}_{QM},0} \right);\left( {0,\underbrace {1, \!\cdots\! ,1}_{{2}QM},0} \right).
	\end{align}}
	{\small \begin{align}\label{ajfokwp3}
	{H_{{A_2}}} \triangleq \left( {0;\underbrace {1, \!\cdots\! ,1}_{QM},\underbrace {0, \!\cdots\! ,0}_{QM + 1}} \right)\left( { - \frac{2}{\alpha };\underbrace {1, \!\cdots\! ,1}_{QM},\underbrace {0, \!\cdots \!,0}_{QM + 1}} \right);\left( {1;\underbrace {0, \cdots ,0}_{QM};\underbrace {1,\! \cdots\! ,1}_{QM},0} \right);\left( { - 1 - \frac{2}{\alpha };\underbrace {0, \cdots ,0}_{QM},\underbrace {1,\! \cdots \!,1}_{QM},0} \right).
	\end{align}}
	{\small \begin{align}\label{aegae1}
	{{\cal A}_{p,q,i}}\!\triangleq\!\frac{{2{A_1}r_q^{2 + {\alpha _\sigma }}}}{{{{e}\ln 2}{\alpha _{{u_e}}}{{\left( {r_2^2 - r_0^2} \right)}}}} H_{4,2:2,1; \cdots ;2,1;0,1}^{0,3:1,2; \cdots ;1,2;1,0}\left(\! {{H_{AS{C_2}}}\left|\!\! {\begin{array}{*{20}{c}}
	{{H_{{A_{1,2}}}}:\left( {1,1} \right),\left( {1 - {m_{{s_1}}},1} \right); \cdots ;\left( {1,1} \right),\left( {1 - {m_{{s_{QM}}}},1} \right); - }\\
	{{H_{{A_{2,2}}}}:\left( {{m_1},1} \right); \cdots ;\left( {{m_{QM}},1} \right);\left( {1,1} \right)}
	\end{array}} \right.}\! \right).
	\end{align}}
	{\small \begin{align}\label{afafarb}
	{H_{{A_{2,2}}}} \triangleq \left( {0;\underbrace {1, \cdots ,1}_{QM},0} \right);\left( {1;\underbrace {1, \cdots ,1}_{QM},0} \right);\left( { - 1 - \frac{2}{\alpha };\underbrace {1, \cdots ,1}_{QM},0} \right).
	\end{align}}
	\setcounter{equation}{\value{mytempeqncnt3}}
	\hrulefill
	\end{figure*}
	\setcounter{equation}{51}
	\begin{IEEEproof}
	Please refer to Appendix \ref{AppendixE}.
	\end{IEEEproof}
	\begin{rem}\label{insightsASR}
	For second communication scenario, the ASR of the RIS-aided path can also be obtained according to Remark \ref{RISshi}. From \eqref{ASRzuizho}, as expected, better channel conditions will result in a higher ASR. In addition, we can also observe that the ASR can be improved by assuring larger RIS because the multivariate Fox's $H$-function in ASR is also more easily affected by $Q$, $M$, and $L$ as SOP and PNSC.
	\end{rem}
	\end{prop}
	\section{Asymptotic Analysis}\label{secapp}
	{\color{blue}
	In the SINR expression \eqref{losusersinr} of scenario 1, we observe that $\left\| {{{\widehat {\bf{h}}}_{\upsilon,\sigma}}} \right\|_F^2$ is the sum of Fisher-Snedecor $ \mathcal{F} $ RVs. An accurate closed-form approximation to the distribution of sum of Fisher-Snedecor $ \mathcal{F} $ RVs using single Fisher-Snedecor $ \mathcal{F} $ distribution has been given in \cite[Theorem 3]{du2019distribution}. Thus, eq. \eqref{losusersinr} can be expressed as		
	\begin{equation}\label{sajfkjeal}
	{\rm{SIN}}{{\rm{R}}_{\upsilon ,\sigma '}} \approx {A_1}{\left( {{d_{\upsilon ,\sigma }}} \right)^{ - {\alpha _\sigma }}}{{\cal F}_1},
	\end{equation}
	where $A_1$ is defined after (22), $ {{\cal F}_1} \sim {\cal F}\left( {{m_{{{\cal F}_1}}},{m_{s{{\cal F}_1}}},{{\bar \gamma }_{{{\cal F}_1}}}} \right) $ is a single Fisher-Snedecor $ \mathcal{F} $ RV which is adopted to approximate $\left\| {{{\widehat {\bf{h}}}_{\upsilon,\sigma}}} \right\|_F^2$, where ${m_{{{\cal F}_1}}},{m_{s{{\cal F}_1}}},$ and ${{\bar \gamma }_{{{\cal F}_1}}}$ can be obtained from the parameters of $\left\| {{{\widehat {\bf{h}}}_{\upsilon,\sigma}}} \right\|_F^2$ with the help of \cite[eq. (14)]{du2019distribution}.
	
In the SINR expression \eqref{SINRlis} of scenario 2, $ {\left\| {{{\widehat {\bf{h}}}_{\upsilon ,{\rm{RIS}}}}} \right\|_F^2} $ is the sum of product of two Fisher-Snedecor $ \mathcal{F} $ RVs. With the help of \cite[eq. (18)]{du2020on} and \cite[eq. (10)]{du2020on}, we can exploit a single Fisher-Snedecor $ \mathcal{F} $ RV to approximate the product of two Fisher-Snedecor $ \mathcal{F} $ RVs. Therefore, the distribution of $ {\left\| {{{\widehat {\bf{h}}}_{\upsilon ,{\rm{RIS}}}}} \right\|_F^2} $ can also be approximated by a single Fisher-Snedecor $ \mathcal{F} $ distribution. Following the similar parameters transformation method as in \eqref{sajfkjeal}, eq. \eqref{SINRlis} can be approximated as
	\begin{equation}\label{sajfkjeal2}
	{\rm{SIN}}{{\rm{R}}_{\upsilon ,{\rm{RIS}'}}}\approx{A_2}{d_{\upsilon ,{\rm{L}}}}^{ - 2}{{\cal F}_2},
	\end{equation}
	where $ {{\cal F}_2} \sim {\cal F}\left( {{m_{{{\cal F}_2}}},{m_{s{{\cal F}_2}}},{{\bar \gamma }_{{{\cal F}_2}}}} \right) $ is a single Fisher-Snedecor $ \mathcal{F} $ RV, and $A_2$ is defined after \eqref{RISdeCDF}.
	
	\begin{them}\label{lastthem}
	For the first communication scenario, let $Z_{\upsilon ,\sigma '}={\rm{SIN}}{{\rm{R}}_{\upsilon ,\sigma '}}$, the CDF of $Z_{\upsilon ,\sigma '}$ can be expressed as
	\begin{equation}\label{aegae1;fk;wo}
	{F_{{Z_{\upsilon ,\sigma '}}}}\left( z \right)= \frac{{2{{\bar \gamma }_{{{\cal F}_1}}}^{{m_{{{\cal F}_1}}}^2 - {m_{{{\cal F}_1}}}}\left( {{G_{1,0}} - {G_{1,2}}} \right)}}{{{\alpha _\sigma }\left( {r_2^2 - r_0^2} \right)\Gamma \left( {{m_{{{\cal F}_1}}}} \right)\Gamma \left( {{m_{s{{\cal F}_1}}}} \right)}},
	\end{equation}
	where $G_{1,\hbar}$ $(\hbar=0,2)$ is derived as
	\begin{equation}\label{wfa;214s}
	G_{1,\hbar}=r_\hbar^2G_{2,0}^{3,0}\left( {\left. {\frac{A_1{\left( {{m_{s{{\cal F}_1}}} - 1} \right)}}{{r_\hbar^{{\alpha _\sigma }}{m_{{{\cal F}_1}}}z{{\bar \gamma }_{{{\cal F}_1}}}^{ - {m_{{{\cal F}_1}}}}}}} \right|\begin{array}{*{20}{c}}
	{1,1 + \frac{{\rm{2}}}{{{\alpha _\sigma }}}}\\
	{0,{m_{s{{\cal F}_1}}},\frac{{\rm{2}}}{{{\alpha _\sigma }}}}
	\end{array}} \right).
	\end{equation}
	\end{them}
	\begin{IEEEproof}
	Please refer to Appendix \ref{last}.
	\end{IEEEproof}
	
	\begin{rem}\label{alkflkre}
		For the second communication scenario, let $Z_{\upsilon ,{\rm{RIS}} '}={\rm{SIN}}{{\rm{R}}_{\upsilon,{\rm{RIS}}'}}$, the CDF of $Z_{\upsilon ,{\rm{RIS}}'}$ can be obtained through simple parameter transformation in \eqref{aegae1;fk;wo}.
			Specifically, we can let $A_1\to A_2$, ${\alpha _\sigma }\to2$, ${m_{{{\cal F}_1}}}\to{m_{{{\cal F}_2}}}$, ${m_{s{{\cal F}_1}}}\to{m_{s{{\cal F}_2}}}$, ${{\bar \gamma }_{{{\cal F}_1}}}\to{{\bar \gamma }_{{{\cal F}_2}}}$, $G_{1,0}\to G_{2,0}$, and $G_{1,2}\to G_{2,2}$.
	\end{rem}
	
Thus, with the help of the approximated CDF of ${\rm{SIN}}{{\rm{R}}_{\upsilon,\sigma'}}$ and ${\rm{SIN}}{{\rm{R}}_{\upsilon,{\rm{RIS}}'}}$, we can re-derive the performance metrics, i.e., OP, SOP, PNSC, and ASR, to simplify the involved calculations. Note that the multivariate Fox's $H$-functions in the CDF expressions of ${\rm{SIN}}{{\rm{R}}_{\upsilon,\sigma}}$ and ${\rm{SIN}}{{\rm{R}}_{\upsilon,{\rm{RIS}}}}$ can be expressed in the form of a multi-fold Mellin-Barnes type contour integration where the order of integration increases with the number of elements in the RIS. Besides, the Meijer $G$-functions in the CDF expressions of ${\rm{SIN}}{{\rm{R}}_{\upsilon,\sigma}}$ and ${\rm{SIN}}{{\rm{R}}_{\upsilon,{\rm{RIS}}}}$ can be written as a one-fold Mellin-Barnes integration. Therefore, the re-derivations of OP, SOP, PNSC, and ASR follow similar methods as Section \uppercase\expandafter{\romannumeral3} A-D, but the results are simpler. In the following, we focus on OP and study the asymptotic behavior in the high-SINR regime, i.e., when $\gamma_{{{\cal F}_1}}\to \infty $, to gain more insights.
	
	For the first communication scenario, the OP can be expressed as
	\begin{equation}\label{afl109}
	OP={\bf {P_A}} {\left[ {{F_{{Z_{{u_n},{\rm L}}}}}(Z_{\rm th}),{F_{{Z_{{u_n},{\rm NL}}}}}(Z_{\rm th})} \right]^{\rm  T}},
	\end{equation}
	which can be evaluated with the aid of the CDF of $Z_{\upsilon ,\sigma '}$.
	To obtain more engineering insights from \eqref{afl109}, the asymptotic behavior in the high-SINR regime is analyzed.
	\begin{prop}\label{lastlastp}
	In the high-SINR regime, the OP can be approximated as
	\begin{equation}\label{aslkdawi-}
	OP={\bf {P_A}} {\left[ {{F_{{Z_{{u_n},{\rm L}'}}}}(Z_{\rm th}),{F_{{Z_{{u_n},{\rm NL}'}}}}(Z_{\rm th})} \right]^{\rm  T}},
	\end{equation}
	where
	{ \begin{equation}\label{afl109213}
	{F_{{Z_{{u_n}\!,\!{\sigma}'}}}}\!=\!\!\frac{{2{{\bar \gamma }_{{\mathcal{F}_1}}}^{ -\! {m_{{\mathcal{F}_1}}}}\!\!\left(\! {r_2^{{2} \!+ \!{\alpha _\sigma } \!{m_{{\mathcal{F}_1}}}}\! \!-\! \! r_0^{{2}\! +\! {\alpha _\sigma }\!{m_{{\mathcal{F}_1}}}}} \!\right)\!{m_{{\mathcal{F}_1}}} \! \!^{{m_{{\mathcal{F}_1}}}\! -\! 1}\!A_1^{m_{{\mathcal{F}_1}}}\!{Z_{\rm th}^{{m_{{\mathcal{F}_1}}}}}}}{{\left( {{2} \!+\! {\alpha _\sigma }{m_{{\mathcal{F}_1}}}} \!\right)\left(\! {r_2^2  \!- \! r_0^2} \right)\!B\!\left(\!{{m_{s{\mathcal{F}_1}}},{m_{{\mathcal{F}_1}}}} \!\right){{\left( {{m_{s{\mathcal{F}_1}}}  \!- \!1} \right)}^{{m_{{\mathcal{F}_1}}}}}}},
	\end{equation}}
	and ${\sigma}'={\rm L}, {\rm NL}$.
	\end{prop}
	\begin{IEEEproof}
	Please refer to Appendix \ref{lastlast}.
	\end{IEEEproof}
	Furthermore, at high-SINRs, the OP can be written as $ OP \propto {{\bar \gamma }_{{\mathcal{F}_1}}}^{ - {G_d}} $, where ${G_d}$ denotes the diversity order. From \eqref{aslkdawi-}, we can observe that ${G_d}=m_{{\mathcal F}_1}$.
	
	As for the second communication scenario, the OP and corresponding asymptotic expression can be obtained through parameter transformation in Remark \ref{alkflkre}.
	}
	
	\section{RIS-aided System With Direct Links}\label{apofpa}
	\begin{figure}[h]
	\centering
	\includegraphics[scale=0.5]{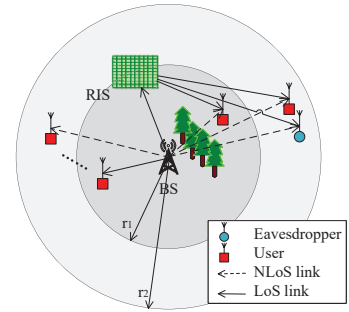}
	\caption{\textcolor{blue}{RIS communication system with the direct links.}}
	\label{model2}
	\end{figure}
	{\color{blue}
	For the second communication scenario, we assume that there is no direct links between the BS and the users when the BS beams toward the RIS. This assumption is reasonable when the signals have high frequency which are easily blocked by obstructions \cite{zhang2021improving}.
	
	However, as shown in Fig. \ref{model2}, a more general SINR expression with direct channels considered can be given as
	\begin{equation}
	{\rm{SIN}}{{\rm{R}}_{\upsilon ,{\rm{RD}}}} = {A_2}d_{\upsilon ,{\rm{L}}}^{ - 2}\left( {\left\| {{{\widehat {\bf{h}}}_{\upsilon ,\sigma }}} \right\|_F^2 + \left\| {{{\widehat {\bf{h}}}_{\upsilon ,{\rm{RIS}}}}} \right\|_F^2} \right).
	\end{equation}
	Note that deriving the exact statistical	characteristics of ${\rm{SIN}}{{\rm{R}}_{\upsilon ,{\rm{RD}}}}$ are challenging, if not impossible, for calculating the performance metrics. Thus, with the help of \eqref{sajfkjeal} and \eqref{sajfkjeal2}, we can obtain the high-quality approximated CDF of ${\rm{SIN}}{{\rm{R}}_{\upsilon ,{\rm{RD}}}}$.
	
	\begin{them}\label{truelast}
	For the second communication scenario with direct links, let $Z_{\upsilon ,RD}={\rm{SIN}}{{\rm{R}}_{\upsilon ,RD}}$, the CDF of $Z_{\upsilon ,RD}$ can be expressed as
	\begin{equation}\label{finaleqrsa}
	F_{Z_{\upsilon ,RD}}(z)\!=\!\prod\limits_{\ell  = 1}^2 \!{{{\left(\! {\frac{{A_2^{ - 1}zr_\hbar ^2{m_{{\mathcal{F}_\ell }}}}}{{\left( {{m_{s{\mathcal{F}_\ell }}} - 1} \right){{\bar \gamma }_{{\mathcal{F}_\ell }}}}}} \!\right)}^{{m_{{\mathcal{F}_\ell }}}}}\!\!\frac{{r_\hbar ^2{{\left( {r_2^2 - r_0^2} \right)}^{ - 1}}G}}{{\Gamma \left( {{m_{s{\mathcal{F}_\ell }}}} \right)\Gamma \left( {{m_{{\mathcal{F}_\ell }}}} \right)}}},
	\end{equation}
	where $G$ is derived as \eqref{s-a=fa=e-faf}, shown at the top of the next page, and $G \, \!\substack{ m_1 , n_1 ; m_2 , n_2 ; m_3 , n_3 \\ p_1 , q_1 ;p_2 , q_2 ; p_3 , q_3}\!(\cdot)\! $ denotes the Bivariate Meijer's $G$-function \cite{shah1973generalizations}.
	\newcounter{mytempeqncnt3HF}
	\begin{figure*}[t]
	\normalsize
	\setcounter{mytempeqncnt3HF}{\value{equation}}
	\setcounter{equation}{60}
	{\small
	\begin{align}\label{s-a=fa=e-faf}
	G \!=\! G_{0,2;3,2;3,2}^{0,1;1,3;1,3}\!\!\left(\!\!\!\! {\left. {\begin{array}{*{20}{c}}
	{\sum\limits_{\ell  = 1}^2 \!{{m_{{\mathcal{F}_\ell }}}}  + 1} \\ 
	{\sum\limits_{\ell  = 1}^2\! {{m_{{\mathcal{F}_\ell }}}} ,2 \!+\!\! \sum\limits_{\ell  = 1}^2 \!{{m_{{\mathcal{F}_\ell }}}} } 
	\end{array}} \!\!\!\right|\!\!\!\!\left. {\begin{array}{*{20}{c}}
	{1 \!-\! {m_{{\mathcal{F}_1}}},1 \!-\! {m_{s{\mathcal{F}_1}}} \!-\! {m_{{\mathcal{F}_1}}},1\! -\! \sum\limits_{\ell  = 1}^2\! {{m_\ell }} } \\ 
	{0, - \sum\limits_{\ell  = 1}^2 \!{{m_{{\mathcal{F}_\ell }} }} } 
	\end{array}}\!\! \right|\!\!\!\!\left. {\begin{array}{*{20}{c}}
	{1\! -\! {m_{{\mathcal{F}_2}}},1\! - \!{m_{s{\mathcal{F}_2}}} \!-\! {m_{{\mathcal{F}_2}}},1 \!- \!\sum\limits_{\ell  = 1}^2\! {{m_{{\mathcal{F}_\ell }} }} } \\ 
	{0, - \sum\limits_{\ell  = 1}^2 {{m_{{\mathcal{F}_\ell }} }} } 
	\end{array}} \!\!\!\!\right|\!\!\!{\begin{array}{*{20}{c}}
	{\frac{{A_2^{ - 1}r_\hbar ^2z{m_{{\mathcal{F}_1}}}}}{{\left( {{m_{s{\mathcal{F}_1}}} - 1} \right){{\bar \gamma }_{{\mathcal{F}_1}}}}}} \\ 
	{\frac{{A_2^{ - 1}r_\hbar ^2z{m_{{\mathcal{F}_2}}}}}{{\left( {{m_{s{\mathcal{F}_2}}} - 1} \right){{\bar \gamma }_{{\mathcal{F}_2}}}}}} 
	\end{array}}} \!\right).
	\end{align}
	}
	\setcounter{equation}{\value{mytempeqncnt3HF}}
	\hrulefill
	\end{figure*}
	\setcounter{equation}{61}
	\end{them}
	\begin{IEEEproof}
	Please refer to Appendix \ref{tryelastA}.
	\end{IEEEproof}
	Therefore, the OP of the second scenario with direct links can be expressed as
	$OP={\bf {P_A}} {\left[ {{F_{{Z_{\upsilon ,L}}}}(Z_{\rm th}),{F_{{Z_{\upsilon ,RD}}}}(Z_{\rm th})} \right]^{\rm  T}}.$
	}
	\section{Numerical Results}\label{sec4}
	In this section, analytical results are presented to illustrate	the advantages of applying the RIS to enhance the security of the DL MIMO communication system. We assume that the noise variances $\sigma_N$ at user $n$ and the eavesdropper are identical and $\sigma_N=0$ ${\rm dB}$. The transmit power ${p_{u_n}}$ is defined in dB with respect to the noise variance. The array gains of main and sidelobes are set to ${G_\upsilon}=30$ ${\rm dB}$ and ${g_\upsilon}=-10$ ${\rm dB}$. For the small scale fading, we set $m_{u_n}=5$, $m_{u_e}=3$, $m_{s_{u_n}}=5$, $m_{s_{u_e}}=3$, ${{\bar{\gamma}}_{u_n}}={{\bar{\gamma}}_{u_e}}=-10$ ${\rm dB}$. The path loss exponent is set to $\alpha_1=2$. Moreover, in the considered simulation scenario, we assume that $r_2 = 400$, $r_1=300$, $r_0=1$, $\beta_{\rm max}=1$, $d_{u_R,{\rm L}}=30$ and $B_1=0.3$. The numerical results are verified via Monte-Carlo simulations by averaging the obtain performance over $10^6$ realizations.
	\begin{figure}[t]
	\centering
	\includegraphics[scale=0.5]{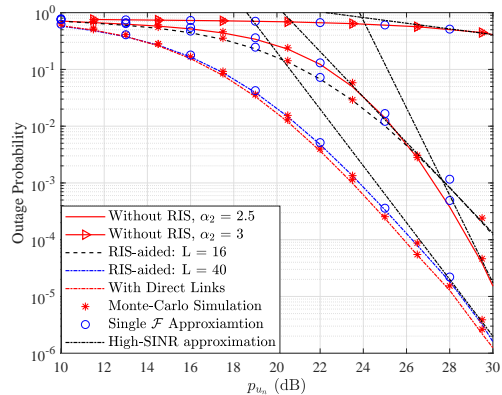}
	\caption{Outage probability versus the transmit power for user $n$.}
	\label{OPFIG}
	\end{figure}
	
	Figure \ref{OPFIG} depicts the OP performance versus transmit power ${p_{u_n}}$	with $K=4$, $M=2$, $Z_{\rm th}=0$ ${\rm dB}$. As it can be observed, the OP decreases as transmit power and $L$ increase, which means that the introduction of a RIS can provide a higher quality of communication for the considered MIMO wireless communication system. {\color{blue}Moreover, when $\alpha_2=2.5$, we can observe that RIS cannot improve the system performance when the number of reflecting elements is small. This is because the OP is dominated by the path loss. When the signal is propagates to the user via the direct links, the signal propagation distance is shorter. Furthermore, while considering the direct links exists between BS and RIS-aided users, we can observe the OP is slightly lower than while assuming no direct links exist. In addition, analytical results perfectly match the approximate ones and Monte-Carlo simulations well. In addition, the asymptotic expressions match well the exact ones at high-SINR values thus proving their validity and versatility.}
	
	\begin{figure}[t]
	\centering
	\includegraphics[scale=0.5]{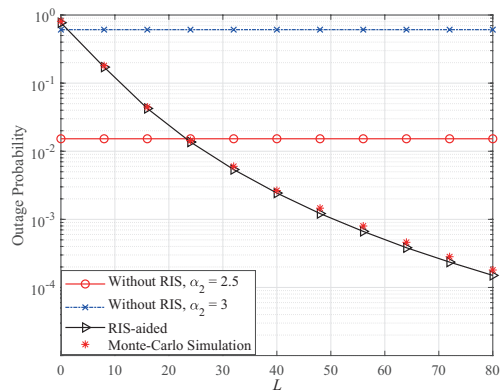}
	\caption{Outage probability versus the number of reflecting elements equipped in RIS.\label{OP2}}
	\end{figure}
	Figure \ref{OP2} illustrates the OP performance versus the number of reflecting elements equipped in RIS, with $K=4$, $M=2$, $Z_{\rm th}=0$ ${\rm dB}$, and $p_{um}=25$ ${\rm dB}$. {\color{blue}In Fig. \ref{OP2}, we can observe that the OP decreases as $L$ increases. This is because the increases of RIS's elements offer more degrees of freedom for efficient beamforming and interference management which can significantly improve the system performance. However, when the path loss of NLoS is small, applying a RIS equipped with a small number of elements, i.e., less than 25, cannot offer a lower OP than the traditional MIMO system. This is because the signal propagation distance in the direct link between the BS and user is shorter then the end-to-end link when RIS is used. In contrast, we can observe that when the path loss of NLoS is large, i.e., when $\alpha_2=3$, the performance gain brought by RIS is significant. Furthermore, when $L$ is sufficiently large such that the reflected signal power by the RIS dominates the total received power at user $n$, we can observe that there is a diminishing return in the slope of OP due to channel hardening \cite{jung2020performance}. Thus, to achieve a low OP at the user, there exists an trade-off between the number of reflecting elements in RIS and the transmit power of the BS.}
	
	\begin{figure}[t]
	\centering
	\includegraphics[scale=0.5]{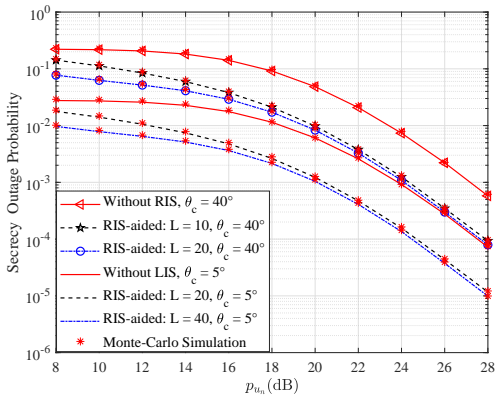}
	\caption{{\color{blue}Secrecy outage probability versus the transmit power for user $n$.}}
	\label{SOPFIG}
	\end{figure}
	Figure \ref{SOPFIG} depicts the SOP performance versus transmit power ${p_{u_n}}$ under different $\theta_c$ with $R_{t}=1$, $K=4$, $M=2$, and $\alpha_2=3$. For a certain setting of the parameters, the SOP decreases as the transmit power of the BS increases. Besides, it can be easily observed that the SOP increases as $\theta_c$ decreases. This is because a large value of $\theta_c$ offers the eavesdroppers a higher possibility in exploiting the larger array gains. Again, it is evident that the analytical results match the Monte-Carlo simulations well.
	
	\begin{figure}[t]
	\centering
	\includegraphics[scale=0.5]{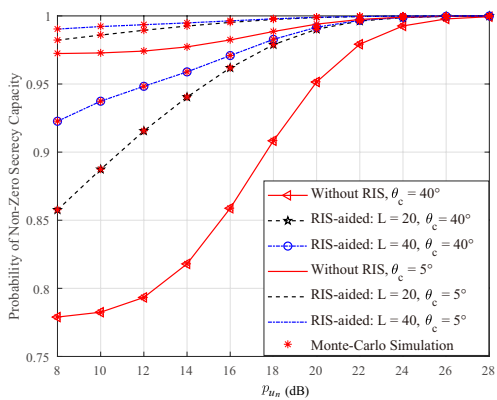}
	\caption{Probability of non-zero secrecy capacity versus transmit power for user $n$.}
	\label{PNSCFIG}
	\end{figure}
	Figure \ref{PNSCFIG} shows the analytical and simulated PNSC versus transmit power ${p_{u_n}}$ under different $\theta_c$ with $R_{t}=1$, $K=4$, $M=2$, and $\alpha_2=3$. As it can be observed, the PNSC decreases as $\theta_c$ increases because the eavesdropper is more likely to receive signals. Moreover, similar to the results in Fig. \ref{SOPFIG}, the use of RIS can enhance the security of the system. {\color{blue}Furthermore, if the transmit signal power is not high or the number of reflecting elements on the RIS is not large, the performance gain brought by RIS is rather then limited, as the signal propagation distance in the direct link between the BS and user is shorter then the end-to-end link when the RIS is used.} However, with the increase of transmit power, the communication scenario with RIS enjoys a better security performance.
	
	\begin{figure}[t]
	\centering
	\includegraphics[scale=0.5]{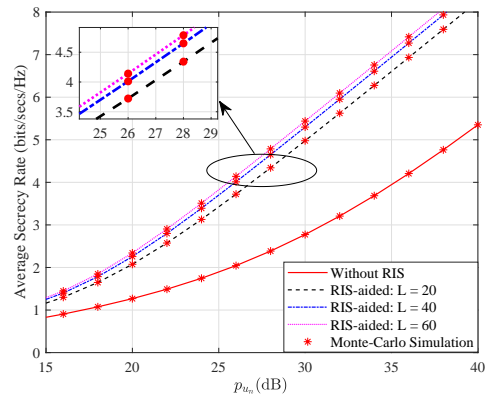}
	\caption{Average security rate versus transmit power for user $n$.}
	\label{ASRFIG}
	\end{figure}
	Figure \ref{ASRFIG} illustrates the ASR as a function of the transmit power ${p_{u_n}}$ for different settings of $L$ with $K=4$, $M=2$, and $\alpha_2=3$. As expected, there is a perfect match between our analytical and simulated results. Like SOP and PNSC, when the transmit power is not large, there is only a small difference for the ASR between communication with RIS and without RIS. This is also because adopting RIS increases the signal propagation distance. In contrast, when the power is moderate to high, RIS is very useful for enhancing the security of the physical layer, and as analyzed in Remark \ref{insightsASR}, the larger the $L$ is, the more obvious the performance gain is. Furthermore, we can observe that there is a diminishing return in the ASR gain due to increasing $L$. This is because when $L$ is large, the security performance of the system is mainly limited by other factors, such as the channel quality and the path loss.
	
	\section{Conclusion}\label{conclusion}
	We proposed a new RIS-aided secure communication system and presented the exact of expressions for CDF and PDF of the SINR for both cases of the RIS is used or not used. Our analytical framework showed the analytical performance expressions of RIS-aided communication subsume the counterpart case without exploiting RIS which can be obtained by some simple parameter transformation of the former case. Secrecy metrics, including the SOP, PNSC, and ASR, were all derived with closed-form expressions in terms of the multivariate Fox's $H$-function. {\color{blue}Numerical results confirmed that RIS can bring significant performance gains and enhance the PLS of the considered system, especially when the number of reflecting elements is sufficiently large and the path loss of NLoS links is severe.}

	\begin{appendices}
	\section{Proof of Theorem \ref{P1}}\label{AppendixA}
	\renewcommand{\theequation}{A-\arabic{equation}}
	\setcounter{equation}{0}
	Based on the result derived in \eqref{losusersinr} and exploiting the fact
	that the elements of ${{{\bf Q}}_{\upsilon,\sigma }}$ are i.n.i.d., the effective channel gain vector of user $n$ (eavesdropper) can be transformed into
	\begin{equation}
	\left\| {{{\widehat {\bf{h}}}_{\upsilon,\sigma}}} \right\|_F^2=\sum\limits_{q = 1}^Q {\sum\limits_{n = 1}^M {{{\left| {{q_{q,n}^{\upsilon,\sigma }}} \right|}^2}} }.
	\end{equation}
	Note that the elements of ${{{\bf Q}}_{\upsilon,\sigma }}$ obey the Fisher-Snedecor $\mathcal{F}$ distribution with fading parameters $m_\ell^{\upsilon,\sigma }$, $m_{s_\ell}^{\upsilon,\sigma }$ and $\Omega_\ell^{\upsilon,\sigma }$. For the sake of presentation, we define $\left\{ {{q_\ell }} \right\}_{i = 1}^{QM} = \left\{ {\left\{ {{q_{q,n}^{\upsilon,\sigma }}} \right\}_{n = 1}^M} \right\}_{q = 1}^Q$.
	
	Letting $\gamma_\ell={\bar\gamma_\ell} {{{\left| {{q_\ell}} \right|}^2}} / \Omega_\ell$, where ${\bar\gamma_\ell}=\mathbb{E}[\Omega_\ell]$, we have
	\begin{equation}
	{f_\gamma }({\gamma _\ell })\! = \!\!\frac{{{m_\ell }^{{m_\ell }} \cdot {{\left( {\left( {{m_{{s_\ell }}} - 1} \right){{\bar \gamma }_\ell }} \right)}^{{m_{{s_\ell }}}}}{\gamma _\ell }^{{m_\ell } - 1}}}{{B\left( {m_\ell,{m_{s_\ell}}} \right){{\left( {{m_\ell }{\gamma _\ell } + \left( {{m_{s_\ell}} - 1} \right){{\bar \gamma }_\ell }} \right)}^{{m_\ell } + {m_{{s_\ell }}}}}}}.
	\end{equation}
	
	Thus, let $X=\sum\limits_{\ell  = 1}^{QM} {{\gamma _\ell }} $, the CDF of $X$ is obtained with the help of \cite[eq. (23)]{rahama2018sum} and \cite[eq (A.1)]{mathai2009h} as
	\begin{align}\label{CDFjifenzhankai}
	{F_X}\left( x \right) &=\prod\limits_{\ell  = 1}^{QM} {\frac{1}{{\Gamma \left( {{m_{{s_\ell }}}} \right)\Gamma \left( {{m_\ell }} \right)}}\int_{{{\cal L}_1}} {\int_{{{\cal L}_2}} { \cdots \int_{{{\cal L}_{_{QM}}}} {{{\left( {\frac{1}{{2\pi j}}} \right)}^{QM}}} } } }\notag\\
	& \times\frac{1}{{\Gamma \left(1+ {\sum\limits_{\ell  = 1}^{QM} {{\zeta _\ell }} } \right)}} \prod\limits_{\ell  = 1}^{QM} {\Gamma \left( {{m_\ell } - {\zeta _\ell }} \right)\Gamma \left( {{\zeta _\ell }} \right)}\notag\\
	&\times\Gamma \!\left( {{m_{{s_\ell }}}\! + \!{\zeta _\ell }} \right) \!\!{\left(\! {\frac{{x{m_\ell }}}{{{{\bar \gamma }_\ell }\left( {{m_{{s_\ell }}} - 1} \right)}}} \!\right)^{  {\zeta _\ell }}}{\rm d}{\zeta _1}\cdots {\rm d}{\zeta _{QM}}.
	\end{align}
	Let $D=\left(d_{u_{m}, \sigma}\right)^{\alpha_{\sigma}}$, hence, using \cite[eq. (30)]{hou2019mimo}, we derive the PDF of $D$ as
	\begin{equation}\label{ddeaerfacifang}
	{f_D}(d) = \frac{2}{{\left( {r_2^2 - r_0^2} \right){\alpha_{\sigma}} }}{d^{\frac{{\rm{2}}}{{\alpha_{\sigma}} } - 1}},\qquad{\rm if}\:\:\:\:{r_0^{\alpha_{\sigma}}  < d < r_2^{\alpha_{\sigma}}}.
	\end{equation}
	Let us define $Y=X/D$, because $X$ and $D$ are statistically independent, the
	CDF of $Y$ can be formulated as
	\begin{equation}\label{SINRdecdfjisuan}
	{F_Y}\!\left( y \right) \!=\!P\left\{ {X \!\le yD} \right\} = \!\int_{r_0^{\alpha_{\sigma}} }^{r_2^{\alpha_{\sigma}} }\!\!\! \!{{F_X}} (xy){f_D}(x){\rm d}x.
	\end{equation}
	Then, by replacing \eqref{CDFjifenzhankai} and \eqref{ddeaerfacifang} into \eqref{SINRdecdfjisuan} and exchanging the order of integration according to Fubini's theorem, we get
	\begin{align}\label{haichayibusinr}
	{F_Y}\left( y \right) = & \frac{2}{{\left( {r_2^2 - r_0^2} \right){\alpha_{\sigma}} }}\prod\limits_{\ell  = 1}^{QM} {\frac{1}{{\Gamma \left( {{m_{st}}} \right)\Gamma \left( {{m_\ell }} \right)}}{{\left( {\frac{1}{{2\pi j}}} \right)}^{QM}}}\notag\\
	\times\!& \int_{{{\cal L}_1}} {\int_{{{\cal L}_2}}  \cdots  } \int_{{{\cal L}_{QM}}} {\frac{1}{{\Gamma \left( {1 + \sum\limits_{\ell  = 1}^{QM} {{\zeta _\ell }} } \right)}}} \notag\\
	\times\!&\prod\limits_{\ell  = 1}^{QM}\! \Gamma \! \left( {{m_\ell }\! - {\zeta _\ell }} \right)\Gamma\! \left( {{\zeta _\ell }} \right)\Gamma \left( {{m_{{s_\ell }}}\! + \!{\zeta _\ell }} \right)\!{\left( \!{\frac{{y{m_\ell }}}{{{{\bar \gamma }_\ell }\left( {{m_{{s_\ell }}} - 1} \right)}}}\! \right)^{{\zeta _\ell }}}\notag\\
	\times\!&\underbrace {\int_{r_0^{\alpha_{\sigma}} }^{r_2^{\alpha_{\sigma}} } {{x^{\sum\limits_{\ell  = 1}^{QM} {{\zeta _\ell }}  + \frac{2}{{\alpha_{\sigma}} } - 1}}} {\rm d}x}_{{I_A}}{\rm d}{\zeta _1}{\rm d}{\zeta _2} \cdots {\rm d}{\zeta _{QM}}.
	\end{align}
	$I_A$ can easily be deduced. Letting $Z=A_1 Y$, we obtain \eqref{FinalSINR1}. The proof is now complete.
	\section{Proof of Theorem \ref{P2}}\label{AppendixB}
	\renewcommand{\theequation}{B-\arabic{equation}}
	\setcounter{equation}{0}
	
	Based on the result derived in \eqref{SINRlis} and exploiting the fact
	that the elements of ${{{\bf Q}}_{3}}$ and ${{{\bf Q}}_{\upsilon,{\rm L}}}$ are i.n.i.d., the effective channel gain vector of user $n$ (eavesdropper) can be transformed into
	\begin{equation}\label{APPBde1}
	\left\| {{{\widehat {\bf{h}}}_{\upsilon,{\rm RIS}}}} \right\|_F^2\!= \!\!\!\sum\limits_{q = 1}^Q\! {\sum\limits_{n = 1}^{L} {{{\left| {\left( {{q_{q,n}^{\upsilon,{\rm L}}}} \right)\left( {{q_{n,m}^{u_R,{\rm L}}}} \right)} \right|}^2}} }  \!= \!\!\sum\limits_{q = 1}^Q\! {\sum\limits_{n = 1}^{L} {{{\left| {{q_{q,n}^{\upsilon,{\rm L}}}} \right|}^2}{{\left| {{q_{n,m}^{u_R,{\rm L}}}} \right|}^2}} } .
	\end{equation}
	Note that ${q^{\upsilon,{\rm L}}_{q,n}} \sim {\cal F}\left( {{m_{q,n}^{\upsilon,{\rm L}}}{,}{m_{{s_{q,n}}}^{\upsilon,{\rm L}}}{,}{\Omega _{q,n}^{\upsilon,{\rm L}}}} \right)$ and ${q_{n,m}^{u_R,{\rm L}}} \sim {\cal F}\left( {{m_{n,m}^{u_R,{\rm L}}}{,}{m_{{s_{n,m}}}^{u_R,{\rm L}}}{,}{\Omega _{n,m}^{u_R,{\rm L}}}} \right)$ are i.n.i.d. Fisher-Snedecor $\mathcal{F}$ RVs. Again, we define that $\left\{ {{q_\ell }} \right\}_{i = 1}^{QL} = \left\{ {\left\{ {\left( {{q_{q,n}^{\upsilon,{\rm L}}}} \right)\left( {{q_{n,m}^{u_R,{\rm L}}}} \right)} \right\}_{n = 1}^{L}} \right\}_{q = 1}^Q$ in order to make our proof process more concise. Thus, using \cite[eq. (18)]{badarneh2018n} and letting $N=2$ and ${\gamma _\ell } = {{\bar \gamma }_\ell }\frac{{{{\left| {{q_\ell }} \right|}^2}}}{{{\Omega _{1,\ell }}{\Omega _{2,\ell }}}}$, the MGF of ${\gamma _\ell }$ can be expressed as
	\begin{align}\label{MGFAPPB4}
	{M_\gamma }\left( {{\gamma _\ell }} \right) =\frac{{G_{3,2}^{2,3}\!\!\left(\!\!\! {\left. {\frac{{{m_{1,\ell }}{m_{2,\ell }}}}{{\left( {{m_{{s_{1,\ell }}}} \!- \!1} \right)\!\left( {{m_{{s_{2,\ell }}}} \!-\! 1} \right){{\bar \gamma }_\ell }{\gamma _\ell }}}} \right|\!\!\!\!\begin{array}{*{20}{c}}
	{1,1 \!-\! {m_{{s_{1,\ell }}}},1 \!- \!{m_{{s_{2,\ell }}}}}\\
	{{m_{1,\ell }},{m_{2,\ell }}}
	\end{array}} \!\!\!\!\right)}}{{\Gamma \left( {{m_{1,\ell }}} \right)\Gamma \left( {{m_{{s_{1,\ell }}}}} \right)\Gamma \left( {{m_{2,\ell }}} \right)\Gamma \left( {{m_{{s_{2,\ell }}}}} \right)}}.
	\end{align}
	Letting $X=\sum\limits_{\ell  = 1}^{QL} {{{\left| {{\gamma _\ell }} \right|}}} $, we can derive the PDF of $X$ as
	\begin{equation}
	f_{X}(x)=\mathcal{L}^{-1}\left\{\mathcal{M}_{X}(s) ; x\right\}=\frac{1}{2 \pi j} \int_{\mathcal{L}} \mathcal{M}_{X}(s) e^{x s} {\rm d}s,
	\end{equation}
	where $\mathcal{L}^{-1}\{\cdot\}$ denotes the inverse Laplace transform. The MGF of the $X$ can be expressed with the aid of \eqref{MGFAPPB4} and \cite[eq. (8.4.3.1)]{gradshteyn2007}
	
	\begin{align}\label{MGFdejifenyangzi}
	f_{X}(x)&=\frac{1}{{2\pi j}}\int_{\cal L} {\prod\limits_{\ell  = 1}^{QL} {{{\cal M}_{{\gamma _\ell }}}\left( s \right){e^{xs}}ds} }\notag\\
	&=\frac{1}{{2\pi j}}\int_{\cal L} {\prod\limits_{\ell  = 1}^{QL} {\frac{{{\left( {{{2\pi j}}} \right)}^{-QL}}}{{\Gamma \left( {{m_{1,\ell }}} \right)\Gamma \left( {{m_{{s_{1,\ell }}}}} \right)\Gamma \left( {{m_{2,\ell }}} \right)\Gamma \left( {{m_{{s_{2,\ell }}}}} \right)}}} } \notag\\
	&\times\!\! \int_{{{\cal L}_{QL}}}\!\!\!\!\!\!\!\cdots\!\! \int_{{{\cal L}_1}}\!\!\!\!{\Upsilon \left( {{\zeta _\ell }} \right)} {\left( {\frac{{{m_{1,\ell }}{m_{2,\ell }}}}{{\left( {{m_{{s_{1,\ell }}}}\! - \!1} \right)\left( {{m_{{s_{2,\ell }}}} \!- \!1} \right){{\bar \gamma }_\ell }s}}} \right)^{ - {\zeta _\ell }}}\notag\\
	&\times {e^{xs}} {\rm d}{\zeta _1} \cdots {\rm d}{\zeta _{QL}}{\rm d}s,
	\end{align}
	where \begin{align}
	\Upsilon \left( {{\zeta _\ell }} \right) = &\Gamma \left( { - {\zeta _\ell }} \right)\Gamma \left( {{m_{{s_{1,\ell }}}} - {\zeta _\ell }} \right)\Gamma \left( {{m_{{s_{2,\ell }}}} - {\zeta _\ell }} \right)\notag\\
	&\times \Gamma \left( {{m_{1,\ell }} + {\zeta _\ell }} \right)\Gamma \left( {{m_{2,\ell }} + {\zeta _\ell }} \right)
	\end{align}
	Note that the order of integration can be interchangeable according to Fubini's theorem, we can re-write \eqref{MGFdejifenyangzi} as
	\begin{align}\label{haichayibuAPPB}
	f_{X}(x)= &\prod\limits_{\ell  = 1}^{QL} {\frac{1}{{\Gamma \left( {{m_{1,\ell }}} \right)\Gamma \left( {{m_{{s_{1,\ell }}}}} \right)\Gamma \left( {{m_{2,\ell }}} \right)\Gamma \left( {{m_{{s_{2,\ell }}}}} \right)}}{{\left( {\frac{1}{{2\pi j}}} \right)}^{QL}}} \notag\\
	&\times \!\int_{{{\cal L}_{QL}}}\!\!\!\!\!\!\cdots \! \int_{{{\cal L}_1}} {\Upsilon \left( {{\zeta _\ell }} \right)} {\left(\! {\frac{{{m_{1,\ell }}{m_{2,\ell }}}}{{\left(\! {{m_{{s_{1,\ell }}}} \!-\! 1} \right)\left( {{m_{{s_{2,\ell }}}} - 1} \right){{\bar \gamma }_\ell }}}} \!\right)^{ - {\zeta _\ell }}}\notag\\
	&\times\underbrace {\frac{1}{{2\pi j}}\int_{\cal L} {{s^{\sum\limits_{\ell  = 1}^{QL} {{\zeta _\ell }} }}} {e^{xs}}{\rm d}s}_{{I_{{\rm B}_1}}}{\rm d}{\zeta _1} \cdots {\rm d}{\zeta _{QL}}.
	\end{align}
	Letting $xs=-t$ and using \cite[eq. (8.315.1)]{gradshteyn2007}, we can derive $I_{{{\rm B}}_1}$ as
	\begin{equation}\label{APPBDEIB}
	I_{{{\rm B}}_1}={{{x^{ - 1 - \sum\limits_{\ell  = 1}^{QL} {{\zeta _\ell }} }}}{\Gamma^{-1} \left( { - \sum\limits_{\ell  = 1}^{QL} {{\zeta _\ell }} } \right)}}.
	\end{equation}
	Substituting \eqref{APPBDEIB} into \eqref{haichayibuAPPB}, we obtain the PDF of $X$, which can be written as Fox's $H$ function. The CDF of $X$ can be expressed as ${F_X}\left( x \right) = \int_0^x {{f_r}\left( r \right)dr}$.
	Thus, we can rewrite the CDF of $X$ as
	\begin{align}\label{CDFhiahai}
	&{F_X}\left( x \right) =\prod\limits_{\ell  = 1}^{QL} {\frac{{{\left( {{{2\pi j}}} \right)}^{-QL}}}{{\Gamma \left( {{m_{1,\ell }}} \right)\Gamma \left( {{m_{{s_{1,\ell }}}}} \right)\Gamma \left( {{m_{2,\ell }}} \right)\Gamma \left( {{m_{{s_{2,\ell }}}}} \right)}}} \notag\\
	&\times\!\!\int_{{{\cal L}_{QL}}} \!\!\!\!\!\! \cdots \! \int_{{{\cal L}_1}} {\frac{{\Upsilon \left( \!{{\zeta _\ell }} \!\right)}}{{\Gamma \left( { - \sum\limits_{\ell  = 1}^L {{\zeta _\ell }} } \right)}}} {\left( {\frac{{{m_{1,\ell }}{m_{2,\ell }}}}{{\left( \!{{m_{{s_{1,\ell }}}} - 1} \right)\left(\! {{m_{{s_{2,\ell }}}} - 1}\! \right){{\bar \gamma }_\ell }}}} \right)^{ - {\zeta _\ell }}}\notag\\
	&\times \underbrace {\int_0^x {{r^{ - 1 - \sum\limits_{\ell  = 1}^{QL} {{\zeta _\ell }} }}{\rm d}r} }_{{I_{{{\rm B}_2}}}}{\rm d}{\zeta _1} \cdots {\rm d}{\zeta _{QL}},
	\end{align}
	where $I_{{\rm B}_2}$ can be solved as
	\begin{equation}\label{IB2}
	I_{{\rm B}_2}=\int_0^x {{r^{ - 1 - \sum\limits_{\ell  = 1}^{QL} {{\zeta _\ell }} }}{\rm d}r}  = \frac{{\rm{1}}}{{^{ - \sum\limits_{\ell  = 1}^{QL} {{\zeta _\ell }} }}}{x^{ - \sum\limits_{\ell  = 1}^{QL} {{\zeta _\ell }} }}.
	\end{equation}
	Substituting \eqref{IB2} into \eqref{CDFhiahai} and using \cite[eq. (8.331.1)]{gradshteyn2007}, equation \eqref{CDFhiahai} can be expressed as
	\begin{align}
	{F_X}\left( x \right) = &\prod\limits_{\ell  = 1}^{QL} {\frac{{{\left( {{{2\pi j}}} \right)}^{-QL}}}{{\Gamma \left( {{m_{1,\ell }}} \right)\Gamma \left( {{m_{{s_{1,\ell }}}}} \right)\Gamma \left( {{m_{2,\ell }}} \right)\Gamma \left( {{m_{{s_{2,\ell }}}}} \right)}}} \notag\\
	&\times \int_{{{\cal L}_{QL}}}\!\!\!\!\!\!\cdots \!\! \int_{{{\cal L}_1}} {\left( {\frac{{x{m_{1,\ell }}{m_{2,\ell }}}}{{\left( {{m_{{s_{1,\ell }}}} - 1} \right)\left( {{m_{{s_{2,\ell }}}} - 1} \right){{\bar \gamma }_\ell }}}} \right)^{ - {\zeta _\ell }}}\notag\\
	&\times \frac{{\Upsilon \left( {{\zeta _\ell }} \right)}}{{\Gamma \left( {1 - \sum\limits_{\ell  = 1}^L {{\zeta _\ell }} } \right)}}{\rm d}{\zeta _1} \cdots {\rm d}{\zeta _{QL}}.
	\end{align}
	\begin{figure}[t]
	\centering
	\includegraphics[scale=0.5]{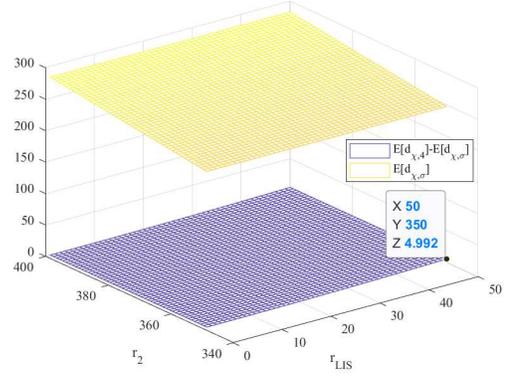}
	\caption{The difference between the expected distance from the user to the BS and the expected distance from the user to the RIS.}
	\label{HPPP}
	\end{figure}
	In order to avoid complex trigonometric operations, we assume that $d_{\upsilon,\sigma}$ and $d_{\upsilon,{\rm L}}$ are i.n.i.d. RVs. To verify that this assumption will only cause small errors for our considered system, we perform Monte-Carlo simulation for the location of user $n$ with $r_0=1$. In the simulation, we generate the location of user $n$ for $10^6$ times according to \cite[eq. (30)]{hou2019mimo}. Fig. \ref{HPPP} shows that when the distance from the RIS to the BS, defined as $r_{{\rm RIS}}$, and the range of the system model, $r_2$, change, the difference between the mean distance from user $n$ to the BS and to the RIS is very small. For example, the maximum is less than $5$ meters.	Thus, let $Y=X/D$, hence, we can derive the CDF of $Y$ using \eqref{SINRdecdfjisuan}
	\begin{align}\label{FYIB3}
	&F_Y(y)\!=\!\frac{2}{{\left( {r_2^2 - r_0^2} \right)\alpha }}\prod\limits_{\ell  = 1}^{QL} {\frac{{{\left( {{{2\pi j}}} \right)}^{-QL}}}{{\Gamma \left( {{m_{1,\ell }}} \right)\!\Gamma \left( {{m_{{s_{1,\ell }}}}} \right)\!\Gamma \left( {{m_{2,\ell }}} \right)\!\Gamma \left( {{m_{{s_{2,\ell }}}}} \right)}}} \notag\\
	&\times\!\!\! \int_{{{\cal L}_{QL}}} \!\!\!\! \cdots\!  \int_{{{\cal L}_1}} {\frac{{\Upsilon \left( \!{{\zeta _\ell }}\! \right)}}{{\Gamma \left( {1 - \sum\limits_{\ell  = 1}^L {{\zeta _\ell }} }\! \right)}}} {\left(\! {\frac{{x{m_{1,\ell }}{m_{2,\ell }}}}{{\left(\! {{m_{{s_{1,\ell }}}} - 1} \right)\left(\! {{m_{{s_{2,\ell }}}} - 1}\! \right){{\bar \gamma }_\ell }}}} \right)^{ - {\zeta _\ell }}}\notag\\
	&\times \underbrace {\int_{{r_0}^\alpha }^{r_2^\alpha } {{y^{\frac{2}{\alpha } - 1 - \sum\limits_{\ell  = 1}^L {{\zeta _\ell }} }}} {\rm d}y}_{{I_{{{\rm B}_3}}}}{\rm d}{\zeta _1} \cdots {\rm d}{\zeta _{QL}}.
	\end{align}
	$I_{{\rm B}_3}$ can be solved easily. Let $\alpha=2$ and $Z=A_2 Y$, we obtain \eqref{RISdeCDF} and complete the proof.
	
	\section{Proof of Proposition \ref{PC}}\label{AppendixC}
	\renewcommand{\theequation}{C-\arabic{equation}}
	\setcounter{equation}{0}
	
	$SOP_{\sigma,{\mathchar'26\mkern-10mu\lambda} }$ can be expressed as
	\begin{equation}\label{chushiSOP}
	SOP_{\sigma,{\mathchar'26\mkern-10mu\lambda} }=\int_0^\infty  {{F_{{Z_{{u_n},\sigma }}}}({R_s}Z + {R_s}- 1){f_{{Z_{{u_e},\sigma,{\mathchar'26\mkern-10mu\lambda}}}}}(Z)} {\rm d}Z.
	\end{equation}
	Substituting \eqref{FinalSINR1} and \eqref{FinalSINR1PDF} into \eqref{chushiSOP}, we obtain
	\begin{align}
	&SOP_{\sigma,{\mathchar'26\mkern-10mu\lambda} }\!= \!\prod\limits_{{\ell _1} = 1}^{QM}\prod\limits_{{\ell _2} = 1}^{QM} \!\!{\frac{{{\Gamma^{-1} \left(\! {{m_{{u_e},{s_{{\ell _2}}}}}} \!\right)\!\Gamma^{-1} \left( \!{{m_{{u_e},{\ell _2}}}} \!\right)}}}{{\Gamma \left(\! {{m_{{u_n},{s_{{\ell _1}}}}}} \!\right)\!\Gamma \left(\! {{m_{{u_n},{\ell _1}}}} \!\right)}}} \notag\\
	&\!\times\!\!\!\!\int_0^\infty \!\!\!\!\!\!\!{\left(\! {{H_{{u_n},CD{F_1},2}}\! -\! {H_{{u_n},CD{F_1},0}}} \!\right)}\!\left(\! {{H_{{u_e},PD{F_1},2}} \!- \!{H_{{u_e},PD{F_1},0}}} \!\right) \!{\rm d}\!Z\notag\\
	&\triangleq{{\cal S}_{2,2}} - {{\cal S}_{2,0}} - {{\cal S}_{0,2}} + {{\cal S}_{0,0}}.
	\end{align}
	Substituting \eqref{H1} and \eqref{H2} into ${{\cal S}_{p,q}}$ and changing the order of integration, the Integral part in ${{\cal S}_{p,q}}$ can be expressed as
	\begin{equation}
	I_1=\int_0^\infty  {{{\left( {{R_s}Z + {R_s} - 1} \right)}^{\sum\limits_{{\ell _1} = 1}^{QM} {{\zeta _{{u_n},{\ell _1}}}} }}{Z^{\sum\limits_{{\ell _2} = 1}^{QM} {{\zeta _{{u_e},{\ell _2}}}} }}{\rm d}Z}.
	\end{equation}
	Let $t = \frac{{{R_s}}}{{{R_s} - 1}}Z$, $I_1$ can be written as
	\begin{align}\label{i1zaizhe}
	I_1= &{\left( {{R_s} - 1} \right)^{\sum\limits_{{\ell _1} = 1}^{QM} {{\zeta _{{u_n},{\ell _1}}}} }}{\left( {\frac{{{R_s} - 1}}{{{R_s}}}} \right)^{{1+}\sum\limits_{{\ell _2} = 1}^{QM} {{\zeta _{{u_e},{\ell _2}}}} }}\notag\\
	&\times\underbrace {\int_0^\infty  {{{\left( {t + 1} \right)}^{\sum\limits_{{\ell _1} = 1}^{QM} {{\zeta _{{u_n},{\ell _1}}}} }}{{\left( t \right)}^{\sum\limits_{{\ell _2} = 1}^{QM} {{\zeta _{{u_e},{\ell _2}}}} }}{\rm d}t} }_{{I_2}},
	\end{align}
	let $\mathcal{L}\{p(t)\}=P(x)$. Using the property of Laplace transform, we have $ \mathcal{L}\left\{\int_{0}^{t} p(z) d z\right\}=\frac{P(x)}{x}$. According to the final value theorem, it follows that
	\begin{equation}\label{zhongzhi}
	\lim _{t \rightarrow \infty}\left(\int_{0}^{t} p(z) d z\right)=e \frac{P(e)}{e}=P(e).
	\end{equation}
	Using \eqref{zhongzhi} and \cite[eq. (2.9.6)]{mathai2009h}, we can solve $I_2$ as
	
	\begin{align}\label{chaojihe}
	{I_2} &\!=\! {\cal L}\left\{\! {{{\left( {t + 1} \right)}^{\sum\limits_{{\ell _1} = 1}^{QM} {{\zeta _{{u_n},{\ell _1}}}} }}{{ t}^{\sum\limits_{{\ell _2} = 1}^{QM} {{\zeta _{{u_e},{\ell _2}}}} }}} \!\right\}\! = \!\Gamma \left(\! {1 \!+\!\! \sum\limits_{{\ell _2} = 1}^{QM} {{\zeta _{{u_e},{\ell _2}}}} } \!\right)\notag\\
	&\times\Psi \left(\! {1\! +\!\! \sum\limits_{{\ell _2} = 1}^{QM} {{\zeta _{{u_e},{\ell _2}}}} ,2 +\!\!\! \sum\limits_{{\ell _1} = 1}^{QM} {{\zeta _{{u_n},{\ell _1}}}}  \!+\! \sum\limits_{{\ell _2} = 1}^{QM} {{\zeta _{{u_e},{\ell _2}}}} ,s} \!\right),
	\end{align}
	where $\Psi(\cdot)$ is the tricomi confluent hypergeometric function \cite[eq. (9.210.2)]{gradshteyn2007}.
	Using \cite[eq. (07.33.07.0003.01)]{web}, \eqref{chaojihe} and \eqref{i1zaizhe} and letting $\left\{ {{t_\ell }} \right\}_{\ell  = 1}^{2QM + 1} = \left\{ {\left\{ {{\zeta _{{u_n},{\ell _1}}}} \right\}_{{\ell _1} = 1}^{QM},\left\{ {{\zeta _{{u_e},{\ell _2}}}} \right\}_{{\ell _2} = 1}^{QM},s} \right\}$, we can derive $S_{2,2}$ as \eqref{SIOsolve} to complete the proof.
	
	\section{Proof of Proposition \ref{PD}}\label{AppendixD}
	\renewcommand{\theequation}{D-\arabic{equation}}
	\setcounter{equation}{0}
	$PNSC_{\sigma,{\mathchar'26\mkern-10mu\lambda} }$ is defined as
	\begin{equation}\label{chushiPNSC}
	PNSC_{\sigma,{\mathchar'26\mkern-10mu\lambda} }= \int_0^\infty  {{F_{{Z_{{u_e},\sigma }}}}} \left( Z \right){f_{{Z_{{u_n},\sigma }}}}\left( Z \right){\rm d}Z.
	\end{equation}
	Substituting \eqref{FinalSINR1} and \eqref{FinalSINR1PDF} into \eqref{chushiPNSC}, we obtain that
	
	\begin{align}
	PNSC_{\sigma,{\mathchar'26\mkern-10mu\lambda} }= \!\!&\prod\limits_{{\ell _1} = 1}^{QM} \!\!{\frac{1}{{\Gamma \!\left(\! {{m_{{u_n},{s_{{\ell _1}}}}}} \!\right)\!\Gamma \left(\! {{m_{{u_n},{\ell _1}}}} \!\right)}}} \!\!\prod\limits_{{\ell _2} = 1}^{QM} \!\!{\frac{1}{{\Gamma \!\left(\! {{m_{{u_e},{s_{{\ell _2}}}}}}\! \right)\!\Gamma \!\left(\! {{m_{{u_e},{\ell _2}}}} \!\right)}}}\notag\\
	&\times\int_0^\infty  {\left( {{H_{{u_n},CD{F_1},2}} - {H_{{u_n},CD{F_1},0}}} \right)}\notag\\
	&\times\left( {{H_{{u_e},PD{F_1},2}} - {H_{{u_e},PD{F_1},0}}} \right) {\rm d}Z\notag\\
	\triangleq&{{\cal P}_{2,2}} - {{\cal P}_{2,0}} - {{\cal P}_{0,2}} + {{\cal P}_{0,0}}.
	\end{align}
	By changing the order of integration, we can express the integral part of ${{\cal P}_{p,q}}$ as
	\begin{equation}\label{I3dechushi}
	{I_3} = \int_0^\infty  {{Z^{\sum\limits_{{\ell _1} = 1}^{QM} {{\zeta _{{u_n},{\ell _1}}}}  + \sum\limits_{{\ell _2} = 1}^{QM} {{\zeta _{{u_e},{\ell _2}}}}  + 1}}{\rm d}Z} ,
	\end{equation}
	which can be solved using the final value theorem as \cite[eq. (34)]{du2019distribution}. Thus, \eqref{PNSCsolve} is obtained, which completes the proof.
	
	\section{Proof of Proposition \ref{PE}}\label{AppendixE}
	\renewcommand{\theequation}{E-\arabic{equation}}
	\setcounter{equation}{0}
	$AS{C_{\sigma ,\lambda }}$ can be expressed as
	\begin{align}\label{chushiASR}
	AS{C_{\sigma ,\lambda }} \!=\!\!& \sum\limits_{i = 1}^2 {{{\cal I}_{\sigma ,\lambda ,i}}} \!- \!{{\cal I}_{\sigma ,\lambda ,3}}
	\!\!=\!\!\sum\limits_{i = 1}^2 \!{\left(\! {{{\cal A}_{2,2,i}} \!\!- \!\!{{\cal A}_{2,0,i}} \!\!-\! \!{{\cal A}_{0,2,i}} \!\!+ \!\!{{\cal A}_{0,0,i}}} \!\right)}\notag\\
	&  - \left( {{{\cal A}_{2,2,3}} - {{\cal A}_{2,0,3}} - {{\cal A}_{0,2,3}} + {{\cal A}_{0,0,3}}} \right),
	\end{align}
	where
	\begin{equation}\label{sanbufen}
	\left\{ \begin{array}{l}\!\!\!
	{{\cal I}_{\sigma ,\mathchar'26\mkern-10mu\lambda ,1}} \!=\! \int_0^\infty  {{{\log }_2}} \left( {1 + Z} \right)\!{f_{{Z_{{u_n},\sigma }}}}\left( Z \right){F_{{Z_{{u_e},\sigma }}}}\left( Z \right){\rm d}Z,\\
	\!\!\!{{\cal I}_{\sigma ,\mathchar'26\mkern-10mu\lambda ,2}} \!=\! \int_0^\infty  {{{\log }_2}} \left( {1 + Z} \right)\!{f_{{Z_{{u_e},\sigma }}}}\left( Z \right){F_{{Z_{{u_n},\sigma }}}}\left( Z \right){\rm d}Z,\\\!\!\!
	{{\cal I}_{\sigma ,\mathchar'26\mkern-10mu\lambda ,3}} \!=\! \int_0^\infty  {{{\log }_2}} \left( {1 + Z} \right)\!{f_{{Z_{{u_e},\sigma }}}}\left( Z \right){\rm d}Z
	\end{array} \right.
	\end{equation}
	and
	\begin{align}
	ASR_{\sigma,{\mathchar'26\mkern-10mu\lambda},i }\!=\! &\prod\limits_{{\ell _1} = 1}^{QM} \!\!{\frac{1}{{\Gamma \!\left(\! {{m_{{u_n},{s_{{\ell _1}}}}}} \!\right)\!\Gamma \!\left(\! {{m_{{u_n},{\ell _1}}}} \!\right)}}}\! \prod\limits_{{\ell _2} = 1}^{QM} {\frac{1}{{\!\Gamma \left(\! {{m_{{u_e},{s_{{\ell _2}}}}}} \!\right)\!\Gamma \!\left(\! {{m_{{u_e},{\ell _2}}}} \!\right)}}}\notag\\
	&\times\int_0^\infty{\log _2}\left( {1 + Z} \right)  {\left( {{H_{{u_n},CD{F_1},2}} - {H_{{u_n},CD{F_1},0}}} \right)} \notag\\
	&\times\left( {{H_{{u_e},PD{F_1},2}} - {H_{{u_e},PD{F_1},0}}} \right){\rm d}Z\notag\\
	\triangleq&{{\cal A}_{2,2,i}} - {{\cal A}_{2,0,i}} - {{\cal A}_{0,2,i}} + {{\cal A}_{0,0,i}}.
	\end{align}
	Substituting \eqref{FinalSINR1} and \eqref{FinalSINR1PDF} into \eqref{sanbufen}, we can express the integral part of ${{\cal I}_{\sigma ,\mathchar'26\mkern-10mu\lambda ,i}}$ $(i=1,2)$ as
	\begin{equation}
	{I_4} = \int_0^\infty  {{\log}_2 \left( {Z + 1} \right){Z^{\sum\limits_{{\ell _1} = 1}^{QM} {{\zeta _{{u_n},{\ell _1}}}}  + \sum\limits_{{\ell _2} = 1}^{QM} {{\zeta _{{u_e},{\ell _2}}}} }}{\rm d}Z},
	\end{equation}
	which has been solved using the final value theorem as \cite[eq. (49)]{du2019distribution}. Thus, ${I_3}$ can be written as
	\begin{align}
	I_4 =&\frac{1}{{{e}\ln 2}} \int_S {\frac{{\Gamma \left( {1 - s} \right){\Gamma ^2}\left( { - \sum\limits_{{\ell _1} = 1}^{QM} {{\zeta _{{u_n},{\ell _1}}}}  + \sum\limits_{{\ell _2} = 1}^{QM} {{\zeta _{{u_e},{\ell _2}}}} } \right)}}{{\Gamma \left( {1 - \sum\limits_{{\ell _1} = 1}^{QM} {{\zeta _{{u_n},{\ell _1}}}}  - \sum\limits_{{\ell _2} = 1}^{QM} {{\zeta _{{u_e},{\ell _2}}}} } \right)}}}\notag\\
	&\times\Gamma \left( {1 + \sum\limits_{{\ell _1} = 1}^{QM} {{\zeta _{{u_n},{\ell _1}}}}  + \sum\limits_{{\ell _2} = 1}^{QM} {{\zeta _{{u_e},{\ell _2}}}} } \right) {e^s}{\rm d}s.
	\end{align}
	With the help of \eqref{sanbufen}, we obtain ${{\cal I}_{\sigma ,\mathchar'26\mkern-10mu\lambda ,i}}$ $(i=1,2)$. Similarly, following the same methodology, one can easily derive ${{\cal I}_{\sigma ,\mathchar'26\mkern-10mu\lambda ,3}}$. The proof is now completed.
	
	\section{Proof of Theorem \ref{lastthem}}\label{last}
	\renewcommand{\theequation}{F-\arabic{equation}}
	\setcounter{equation}{0}
	{\color{blue}The CDF of $\mathcal{F}_1$ is given as \cite[eq. 12]{yoo2019a}
	\begin{align}\label{safk12}
	&F_{{{\cal F}_1}}\left( {{{\cal F}_1}} \right) = \frac{{{m_{{{\cal F}_1}}}^{{m_{{{\cal F}_1}}} - 1}{{\cal F}_1}^{{m_{{{\cal F}_1}}}}}}{{B\left( {{m_{{{\cal F}_1}}},{m_{s{{\cal F}_1}}}} \right){{\left( {{m_{s{{\cal F}_1}}} - 1} \right)}^{{m_{{{\cal F}_1}}}}}{{\bar \gamma }_{{{\cal F}_1}}}^{{m_{{{\cal F}_1}}}}}}
	\notag\\&\times{}_2{F_1}\!\left( {{m_{{{\cal F}_1}}},{m_{{{\cal F}_1}}} + {m_{s{{\cal F}_1}}},{m_{{{\cal F}_1}}} \!+ 1; \frac{{- {m_{{{\cal F}_1}}}{{\cal F}_1}}}{{\left( {{m_{s{{\cal F}_1}}} - 1} \right){{\bar \gamma }_{{{\cal F}_1}}}^{{m_{{{\cal F}_1}}}}}}} \right).
	\end{align}
	Let us define $Y={\mathcal{F}_1}/D$, the CDF of $Y$ can be formulated as
	\begin{equation}\label{lk90912}
	{F_Y}\left( y \right) = P\left\{ {{{\cal F}_1} \le yD} \right\} = \int_{r_0^{{\alpha _\sigma }}}^{r_2^{{\alpha _\sigma }}}  {F_{{{\cal F}_1}}}({{\cal F}_1}y){f_D}({{\cal F}_1}){\rm{d}}{{\cal F}_1}.
	\end{equation}
	Substituting \eqref{safk12} and \eqref{ddeaerfacifang} into \eqref{lk90912} and using \cite[eq. (9.113)]{gradshteyn2007}, we obtain
	\begin{align}
	&{F_Y}\left( y \right) = \frac{{{m_{{{\cal F}_1}}}^{{m_{{{\cal F}_1}}} - 1}{y^{{m_{{{\cal F}_1}}}}}{{\bar \gamma }_{{{\cal F}_1}}}^{ - {m_{{{\cal F}_1}}}}}}{{{\alpha _\sigma }\left( {r_2^2 - r_0^2} \right)B\left( {{m_{{{\cal F}_1}}},{m_{s{{\cal F}_1}}}} \right){{\left( {{m_{s{{\cal F}_1}}} - 1} \right)}^{{m_{{{\cal F}_1}}}}}}}
	\notag\\&
	\times \frac{{\Gamma \left( {{m_{{{\cal F}_1}}} + 1} \right)}}{{\Gamma \left( {{m_{{{\cal F}_1}}}} \right)\Gamma \left( {{m_{{{\cal F}_1}}} + {m_{s{{\cal F}_1}}}} \right)}}\frac{1}{{\pi j}}
	\notag\\&
	\times\int_{\cal L} {\frac{{\Gamma \left( {{m_{{{\cal F}_1}}} + t} \right)\Gamma \left( {{m_{{{\cal F}_1}}} + {m_{s{{\cal F}_1}}} + t} \right)}}{{\Gamma \left( {{m_{{{\cal F}_1}}} + 1 + t} \right)}}} {\left( {\frac{{{m_{{{\cal F}_1}}}y{{\bar \gamma }_{{{\cal F}_1}}}^{ - {m_{{{\cal F}_1}}}}}}{{\left( {{m_{s{{\cal F}_1}}} - 1} \right)}}} \right)^t}
	\notag\\&
	\times\underbrace {\int_{r_0^{{\alpha _\sigma }}}^{r_2^{{\alpha _\sigma }}}  {{\cal F}_1}^{t + {m_{{{\cal F}_1}}} + \frac{{\rm{2}}}{{{\alpha _\sigma }}} - 1}{\rm{d}}{{\cal F}_1}}_{{I_{{F_1}}}}{\rm{d}}t,
	\end{align}
	where the integration path of $\mathcal{L}$ goes from $t-\inf j$ to $t+\inf j$ and $t\in\mathbb{R}$. $I_{F_1}$ can be solved easily. With the help of \cite[eq. (9.301)]{gradshteyn2007}, \cite[eq. (9.31.5)]{gradshteyn2007}, \cite[eq. (9.113)]{gradshteyn2007}, \cite[eq. (8.384.1)]{gradshteyn2007} and \cite[eq. (8.331.1)]{gradshteyn2007}, ${F_Y}\left( y \right)$ can be expressed by the Meijer's $G$-function. Let $ {Z_{\upsilon ,\sigma '}} = {A_1}Y $, we obtain \eqref{aegae1;fk;wo} that complete the proof.
	}
	\section{Proof of Proposition \ref{lastlastp}}\label{lastlast}
	\renewcommand{\theequation}{G-\arabic{equation}}
	\setcounter{equation}{0}
	{\color{blue}
	With the help of \cite[eq. (9.301)]{gradshteyn2007}, eq. \eqref{wfa;214s} can be expressed as \eqref{213/;}, shown at the top of the next page.
	\newcounter{mytempeqncnt3G}
	\begin{figure*}[t]
	\normalsize
	\setcounter{mytempeqncnt3G}{\value{equation}}
	\setcounter{equation}{0}
	\begin{equation}\label{213/;}
	G_{2,0}^{3,0}\left( {\left. {\frac{A_1{\left( {{m_{s{{\cal F}_1}}} - 1} \right)}}{{r_\hbar^{{\alpha _\sigma }}{m_{{{\cal F}_1}}}z{{\bar \gamma }_{{{\cal F}_1}}}^{ - {m_{{{\cal F}_1}}}}}}} \right|\begin{array}{*{20}{c}}
	{1,1 + \frac{{\rm{2}}}{{{\alpha _\sigma }}}}\\
	{0,{m_{s{{\cal F}_1}}},\frac{{\rm{2}}}{{{\alpha _\sigma }}}}
	\end{array}} \right) = \frac{1}{{2\pi j}}\int_{\cal L} {{{\left( {\frac{A_1{\left( {{m_{s{{\cal F}_1}}} - 1} \right)}}{{r_\hbar^{{\alpha _\sigma }}{m_{{{\cal F}_1}}}z{{\bar \gamma }_{{{\cal F}_1}}}^{ - {m_{{{\cal F}_1}}}}}}} \right)}^s}} \frac{{\Gamma \left( { - s} \right)\Gamma \left( {{m_{s{{\cal F}_1}}} - s} \right)\Gamma \left( {\frac{{\rm{2}}}{{{\alpha _\sigma }}} - s} \right)}}{{\Gamma \left( {1 - s} \right)\Gamma \left( {1 + \frac{{\rm{2}}}{{{\alpha _\sigma }}} - s} \right)}}ds.
	\end{equation}
	\setcounter{equation}{\value{mytempeqncnt3G}}
	\hrulefill
	\end{figure*}
	\setcounter{equation}{1}
	When $\gamma_{\mathcal{F}_1} \to \infty$, we have ${\frac{{\left( {{m_{s{{\cal F}_1}}} - 1} \right)}}{{r_2^{{\alpha _\sigma }}{m_{{{\cal F}_1}}}z{{\bar \gamma }_{{{\cal F}_1}}}^{ - {m_{{{\cal F}_1}}}}}}}\to\infty$. The Mellin-Barnes integral over $\mathcal{L}$ can be calculated approximately by evaluating the residue at $-m_{\cal{F}}$ according to \cite[Theorem 1.7]{mathai2009h}. Thus, we obtain
	\begin{align}\label{alks-0a=ef90}
	&G_{2,0}^{3,0}\left( {\left. {\frac{A_1{\left( {{m_{s{{\cal F}_1}}} - 1} \right)}}{{r_\hbar^{{\alpha _\sigma }}{m_{{{\cal F}_1}}}z{{\bar \gamma }_{{{\cal F}_1}}}^{ - {m_{{{\cal F}_1}}}}}}} \right|\begin{array}{*{20}{c}}
	{1,1 + \frac{{\rm{2}}}{{{\alpha _\sigma }}}}\\
	{0,{m_{s{{\cal F}_1}}},\frac{{\rm{2}}}{{{\alpha _\sigma }}}}
	\end{array}} \right)
	\notag\\&\underrightarrow {{{\bar \gamma }_{{\mathcal{F}_1}}} \to \infty }
	\frac{{\Gamma \left( {{m_{s{\mathcal{F}_1}}} + {m_{{\mathcal{F}_1}}}} \right)}}{{{m_{{\mathcal{F}_1}}}\left( {\frac{{\text{2}}}{{{\alpha _\sigma }}} + {m_{{\mathcal{F}_1}}}} \right)}}{\left( {\frac{{r_2^{{\alpha _\sigma }}{m_{{\mathcal{F}_1}}}y}}{{{{\bar \gamma }_{{\mathcal{F}_1}}}^{{m_{{\mathcal{F}_1}}}}\left( {{m_{s{\mathcal{F}_1}}} - 1} \right)}}} \right)^{{m_{{\mathcal{F}_1}}}}}.
	\end{align}
	Substituting \eqref{alks-0a=ef90} into \eqref{aegae1;fk;wo}, after some algebraic manipulations, we obtain \eqref{aslkdawi-} and complete the proof.
	}
	\section{Proof of Theorem \ref{truelast}}\label{tryelastA}
	\renewcommand{\theequation}{H-\arabic{equation}}
	\setcounter{equation}{0}
	{\color{blue}
	With the help of \eqref{sajfkjeal} and \eqref{sajfkjeal2} , the CDF of $Z_{\upsilon ,RD}$ can be regarded as the CDF of sum of two Fisher $\mathcal{F}$ RVs. Letting $X=\mathcal{F}_1+\mathcal{F}_2$, using \cite[eq. (12)]{du2019distribution} and \cite{mathai2009h}, we can derive the CDF of $X$ as \eqref{s-a=fa=e-f}, shown at the top of the next page.
	\newcounter{mytempeqncnt3H}
	\begin{figure*}[t]
	\normalsize
	\setcounter{mytempeqncnt3H}{\value{equation}}
	\setcounter{equation}{0}
	\begin{align}\label{s-a=fa=e-f}
	{F_X}\left( x \right) =& {\prod\limits_{\ell  = 1}^2 {\left( {\frac{{x{m_{{\mathcal{F}_\ell }}}}}{{\left( {{m_{s{\mathcal{F}_\ell }}} - 1} \right){{\bar \gamma }_{{\mathcal{F}_\ell }}}}}} \right)} ^{{m_{{\mathcal{F}_\ell }}}}}{\left( {\frac{1}{{2\pi j}}} \right)^2}\int_{{\mathcal{L}_2}} {\int_{{\mathcal{L}_1}} {\frac{1}{{\Gamma \left( {\sum\limits_{\ell  = 1}^2 {\left( {{m_{{\mathcal{F}_\ell }}} + {s_\ell }} \right)} } \right)}}} } 
	\notag\\&\times
	\prod\limits_{\ell  = 1}^2 {\frac{{\Gamma \left( {{m_{{\mathcal{F}_\ell }}} + {s_\ell }} \right)\Gamma \left( {{m_{s{\mathcal{F}_\ell }}} + {m_{{\mathcal{F}_\ell }}} + {s_\ell }} \right)\Gamma \left( { - {s_\ell }} \right)\Gamma \left( {{s_\ell } + \sum\limits_{\ell  = 1}^2 {{m_{{\mathcal{F}_\ell }} }} } \right)}}{{\Gamma \left( {{m_{s{\mathcal{F}_\ell }}}} \right)\Gamma \left( {{m_{{\mathcal{F}_\ell }}}} \right)\Gamma \left( {{s_\ell } + \sum\limits_{\ell  = 1}^2 {{m_{{\mathcal{F}_\ell }} }}  + 1} \right)}}} {\left( {\frac{{x{m_{{\mathcal{F}_\ell }}}}}{{\left( {{m_{s{\mathcal{F}_\ell }}} - 1} \right){{\bar \gamma }_{{\mathcal{F}_\ell }}}}}} \right)^{{s_\ell }}}d{s_1}d{s_2}.
	\end{align}
	\setcounter{equation}{\value{mytempeqncnt3H}}
	\hrulefill
	\end{figure*}
	\setcounter{equation}{1}
	
	Let $Y=X/D$, hence, we can derive the CDF of $Y$ with the help of  \eqref{SINRdecdfjisuan}. Using $Z=A_2Y$, we can obtain the CDF of $Z$ in terms of Bivariate Meijer's $G$-function as \eqref{finaleqrsa} to complete the proof.

	}
	\end{appendices}
	\bibliographystyle{IEEEtran}
	\bibliography{IEEEabrv,Ref}
	\end{document}